%% file: main.tex
\documentclass[twoside,11pt]{article}
\usepackage{jair, theapa, rawfonts}

\usepackage[utf8]{inputenc}
\usepackage{amsmath}
\usepackage{amsfonts}
\usepackage{amssymb}
\usepackage{graphicx}
\usepackage{dsfont}

\usepackage{xcolor}
\usepackage{colortbl}
\usepackage{booktabs}
\usepackage{subcaption}
\usepackage{textcomp}
\usepackage[lined,boxed,commentsnumbered]{algorithm2e}
\usepackage{url}
\usepackage[normalem]{ulem}
\usepackage{enumitem}
\setlist[itemize]{leftmargin=*}
\usepackage{mathtools}

\newcommand{\Simplex}{\Delta^{n-1}}
\DeclareMathOperator*{\argmin}{arg\,min}
\newcommand{\ind}[1]{ \mathds{1} {#1}}

\newcommand{\x}{\mathbf{x}}
\newcommand{\namedpar}[1]{\vspace{0cm} \textbf{#1}}

\newif\ifdraft
\newif\ifrevision

\newcommand{\alexcomment}[1]{\ifdraft{\leavevmode\color{violet}{[AM]: {#1}}}\else{\vspace{0ex}}\fi}

\newcommand{\afabcomment}[1]{\ifdraft{\leavevmode\color{blue}{[AF]: {#1}}}\else{\vspace{0ex}}\fi}

\usepackage[show]{chato-notes}

\newcommand{\remove}[1]{\ifrevision{\textcolor{magenta}{\sout{#1}}}\else{}\fi}
\newcommand{\replace}[2]{\ifrevision{\textcolor{magenta}{\sout{#1}}\textcolor{magenta}{{#2}}}\else{#2}\fi}
\newcommand{\add}[1]{\ifrevision{\textcolor{magenta}{#1}}\else{#1}\fi}

\usepackage{adjustbox}
\usepackage{makecell}
\usepackage[noadjust]{marginnote}

\ShortHeadings{Quantifying Query Fairness Under Unawareness}{Jaenich, Moreo, Fabris, McDonald, Esuli, Ounis \& Sebastiani}
\firstpageno{1}

\begin{document}

\title{Quantifying Query Fairness Under Unawareness}

\author{\name Thomas Jaenich* \\ \email t.jaenich.1@research.gla.ac.uk\\
\addr University of Glasgow \\ Glasgow, UK
\AND
       \name Alejandro Moreo*\\ \email alejandro.moreo@isti.cnr.it \\
       \addr Istituto di Scienza e Tecnologie dell'Informazione, Consiglio Nazionale delle Ricerche
       \\ 56124 Pisa, Italy
       \AND
       \name Alessandro Fabris\\ \email alessandro.fabris@mpi-sp.org\\
       \addr Max Planck Institute for Security and Privacy (MPI-SP)
       \\
      Bochum, Germany
       \AND
       \name Graham McDonald \\ \email Graham.Mcdonald@glasgow.ac.uk\\  \addr University of Glasgow \\ Glasgow, UK 
        \AND
         \name Andrea Esuli \\ \email andrea.esuli@isti.cnr.it \\ \addr Istituto di Scienza e Tecnologie dell'Informazione, Consiglio Nazionale delle Ricerche \\ 56124 Pisa, Italy
          \AND
         \name Iadh Ounis\\ \email Iadh.Ounis@glasgow.ac.uk \\ \addr University of Glasgow\\ Glasgow, UK
          \AND
         \name Fabrizio Sebastiani \\ \email fabrizio.sebastiani@isti.cnr.it \\ \addr Istituto di Scienza e Tecnologie dell'Informazione, Consiglio Nazionale delle Ricerche \\ 56124 Pisa, Italy
         }


\maketitle

\begin{abstract}
\noindent \replace{The primary objective of traditional ranking algorithms is to retrieve items that are most relevant to a user's query. However, undesirable biases learnt from the data can skew algorithms and decisions against vulnerable groups of stakeholders. Typically, unfairness in an information access system (IAS) is measured by comparing the distribution of groups in a ranking with a target distribution, such as the distribution of groups in the entire collection. These metrics rely on the true group labels for each item in the ranking. When groups are defined using demographic or sensitive information, these labels are often unknown, a situation referred to as "fairness under unawareness."
To determine the group membership of individual items, labels can be inferred using machine-learned classifiers, with group prevalence estimated by counting the predicted labels. However, these methods often perform poorly under dataset shifts, resulting in inaccurate fairness assessments. In this paper, we propose a robust fairness estimator based on quantification. Our method, which extends beyond binary sensitive groups, significantly outperforms existing baselines for multiple sensitive attributes. To our knowledge, this is the first work to establish a robust protocol for reliably measuring fairness under unawareness across multiple groups of items.}
{Traditional ranking algorithms are designed to retrieve the most relevant items for a user’s query, but they often inherit biases from data that can unfairly disadvantage vulnerable groups. Fairness in information access systems (IAS) is typically assessed by comparing the distribution of groups in a ranking to a target distribution, such as the overall group distribution in the dataset. These fairness metrics depend on knowing the true group labels for each item. However, when groups are defined by demographic or sensitive attributes, these labels are often unknown, leading to a setting known as ``fairness under unawareness.'' To address this, group membership can be inferred using machine-learned classifiers, and group prevalence is estimated by counting the predicted labels. Unfortunately, 
such an estimation is known to be unreliable under dataset shift, compromising the accuracy of fairness evaluations. In this paper, we introduce a robust fairness estimator based on quantification that effectively handles multiple sensitive attributes beyond binary classifications. Our method outperforms existing baselines across various sensitive attributes and, to the best of our knowledge, is the first to establish a reliable protocol for measuring fairness under unawareness across multiple queries and groups.}
\end{abstract}

\section{Introduction}
\label{sec:intro}
\noindent 
In addition to ensuring the relevance of search results, preventing unfairness and discrimination in ranking has become a fundamental objective in the development of \replace{Information Retrieval (IR) systems}{information access systems (IAS)}~\cite{ekstrand2022fairness,zehlike2022fairness}. With this in mind, there have been many approaches proposed in the literature for mitigating unfairness in the results of \replace{IR systems}{IAS}~\cite{biega2018equity,geyik2019fairness,heuss2022fairness,jaenich2023colbert,jaenich2024fairness,morik2020controlling,singh2018fairness}. Providing fair search results is crucial, since ranking \replace{documents}{items} that belong to groups identified by sensitive attributes can significantly impact real-world outcomes, such as economic opportunities~\cite{chen2018investigating,pedreshi2008discrimination}. When \replace{documents}{items} that are associated with a specific demographic attribute are systematically ranked lower in the search results than \replace{documents}{items}  from other demographics, the low-ranked \replace{documents}{items}  will receive less attention from users, since items high up in the ranking are more likely to be examined by users~\cite{craswell2008experimental}. This can be problematic in practical scenarios. For example, in job search, this positional bias means that recruiters may only notice candidates from the top-ranked applications, potentially overlooking qualified individuals who are ranked lower. \add{While this will not always warrant a bias-mitigating intervention, it is imperative to at least \emph{evaluate} and \emph{monitor} the fairness of rankings in high-stakes domain~\cite{EUAIACT2024,NYC2021LL144}}.


Another important factor influencing the fairness of search results is the query that is issued by the user. The degree of unfairness in the search results can vary across different queries, depending on how a query is formulated. Such unfairness can be introduced either directly by the user, e.g., through the replication of existing biases when formulating the query~\cite{Kopeinik2023MaleNurse}, or automatically, e.g., through the auto-completion features of a search system~\cite{chen2018investigating}. Therefore, assessing the fairness of search results related to a specific query is crucial to determine whether fairness interventions are needed. We introduce the terminology \textit{query fairness estimation} (QFE) for the task of assessing the fairness of search results for a given query.

To perform QFE, one typically needs access to the group labels of the ranked \replace{documents}{items}~\cite{kuhlman2021measuring,raj2022measuring,zehlike2022fairness}. These labels categorise \replace{documents}{items} by sensitive attributes, such as race or gender. However, access to these group labels is often limited due to legal, ethical, or other data availability constraints~\cite{bogen2020awareness,holstein2019improving}. As a result, fairness evaluations must often occur under ``unawareness,'' where the labels are unknown.

One way to obtain the labels under unawareness is to use human annotators. However, this is costly and impractical in most scenarios. An alternative is to deploy classifiers to infer the document labels automatically. A classification method deployed in practice is the \textit{Bayesian improved surname geocoding} (BISG) that is used to infer race from surnames and ZIP codes using Bayesian statistics and US Census data~\cite{adjaye2014using}. While cost-efficient, deploying classifiers can introduce more unintended unfairness and bias by the classifiers themselves~\cite{ghosh2021fair}. 

Moreover, even seemingly accurate classifiers can lead to unreliable results when performing QFE. For example, a good classifier may achieve high accuracy by focusing disproportionately on one class, thus minimising errors like false positives at the expense of increasing errors like false negatives \cite<see \S 1.2 in>{Esuli2023}. While technically accurate, this skewed performance fails to reflect the true distribution of groups, making it unreliable for assessing the proportions in a broader set of documents, and can lead to significant errors and misjudgments in the measurements of fairness~\cite{chen2019fairness}. 


\replace{To overcome the aforementioned shortcomings of using standard classification, in}{In} this work, we propose the use of quantification techniques, i.e., machine learning models specifically trained to estimate the relative frequencies of the classes in unlabelled data~\cite{Esuli2023} to improve QFE. Specifically, our main focus is QFE under unawareness of sensitive attributes with multiple classes. 
To the best of our knowledge, our proposed approach is the first to cover groups with non-binary protected attributes and to estimate ranking fairness across multiple queries. 
Our main contributions are as follows: 
\begin{itemize}
    \item We propose a new family of principled methods to perform Query Fairness Estimation across multiple queries and non-binary sensitive attributes.
    \item  We introduce the first approach designed to make quantification methods robust against sample selection bias.
    \item Through extensive experiments on the TREC 2022 Fair Ranking Track collection~\cite{trec-fair-ranking-2022}, we demonstrate that our quantification-based approach outperforms previous methods.
\end{itemize}



\section{Related Work}
\label{sec:related}
\noindent In recent years, ensuring fairness in search results has emerged as a crucial objective alongside the traditional goal of relevance in the development of \replace{IR systems}{IAS}~\cite{ekstrand2022fairness,zehlike2022fairness}. To measure the fairness of search results for a given query, i.e., for the task of QFE, several measures have been introduced~\cite{biega2018equity,diaz2020evaluating,kirnap2021estimation,kuhlman2021measuring,raj2022measuring,sapiezynski2019quantifying,singh2018fairness,yang2024language,yang2017measuring}. While these measures cover different notions of fairness, for example \textit{equality of opportunity}~\cite{biega2018equity,hardt2016equality,diaz2020evaluating,singh2018fairness} or \textit{statistical parity}~\cite{geyik2019fairness,sapiezynski2019quantifying,zehlike2017fa}, they all depend on accurate knowledge of the document labels in a ranking. 
In this work, we consider an ``unawareness'' scenario, where group labels are unavailable, requiring alternative solutions to ensure accurate fairness assessments.

Related to this, \cite{ghosh2021fair} have conducted an extensive study showing that inferring labels using standard classifiers can be problematic. Their work highlights the need for reliable methods to access accurate fairness labels. Although several studies have focused on enhancing the fairness of classifiers with noisy or incomplete group labels~\cite{celis2021fair,friedler2021possibility,ghosh2023fair,mozannar2020fair,wang2020robust}, they primarily address fairness metrics for classification, not for ranking tasks.

In this work, we focus specifically on QFE for rankings under unawareness, an area that has received less attention compared to classification. In the absence of group labels, \cite{chen2023learn} proposed a distribution-based learning approach that leverages contextual features. Their approach uses a loss function that does not require explicit group labels but instead targets a fair distribution. Unlike their task of mitigating unfairness, our main objective is to obtain reliable estimates of group proportions in a ranking to accurately measure the fairness of search results.

\cite{kirnap2021estimation} proposed a method using a small query-dependent subset of data annotated by human assessors for QFE. Moreover, they only focus on binary group fairness, comparing protected versus non-protected groups. Our work addresses the characteristics of multiclass groups and does not require impractical human annotations for each query.

Related to our work on QFE in rankings, \cite{ghazimatin2022measuring} have proposed an approach that we term Post-Metric Correction (PMC). Both our approach and the PMC variants include a correction phase and rely on the outputs of an underlying classifier. However, there are significant differences between our quantification-based approaches and the PMC variants. First, the PMC variants are designed only for binary group fairness assessment; our method natively caters to multi-valued sensitive attributes. This is exceedingly rare and important in algorithmic fairness research \cite{fabris2024lazy}. Moreover, each PMC variant is tied to a specific independence assumption, which may prove difficult to verify in general settings. Finally, the PMC methods are also tailored to one specific fairness metric, while our approach is more versatile. Our method applies a general pre-correction to the class prevalence estimates, which are then used to compute different fair ranking metrics.

In our work, we propose to use quantification methods to make QFE robust to the limitations when standard classifiers are applied. In a related effort, \cite{fabris2023measuring} have shown that using quantification techniques is a useful way to assess the fairness of algorithms when the labels are unknown. However, their work focuses on classification problems, while our work focuses specifically on QFE.


\section{Proposed Approach}
\label{sec:method}
\subsection{Running Example}
Consider an online hiring platform assisting recruiters to fill job openings with promising candidates. Recruiters query the platform with job descriptions and get ranked lists of candidates in return, in decreasing order of estimated job fitness. Platform developers want to ensure that their models are fair with respect to multiple attributes, including age, gender, race, and ethnicity. Given the sensitive nature of this information, most data subjects will be hesitant to disclose it \cite{bogen2021:ap}. Developers, therefore, obtain sensitive attributes for a subset of users through voluntary data disclosure \cite{wilson2021building,linkedin2024settings}. This incomplete demographic data can be used to estimate platform fairness across all queries and users. Finding the optimal way to perform this estimate is an open research problem tackled in the remainder of this section.

\subsection{Learning to Quantify}

\noindent The field of quantification emerges from the fundamental observation that counting over the labels predicted by a classifier tends to produce poor estimates of class prevalence~\cite<see \S 1.2 in>{Esuli2023}, unless the classifier is a perfect one. The above na\"ive counting approach has come to be known as the ``Classify \& Count'' (CC) method~\cite{Forman:2005fk}, and nowadays represents the strawman baseline any proper quantification method is expected to beat. 

More formally, a quantifier is a function $\lambda : \mathbb{N}^\mathcal{X} \rightarrow \Simplex$ mapping bags (or multi-sets) of instances from the input space $\mathcal{X}=\mathbb{R}^d$ to the probability simplex, so that $\lambda(\mathbf{X})=\mathbf{p}$ lies on the $(n-1)$-simplex defined as $\Simplex=\{p_1, \ldots, p_n : p_i \geq 0, \sum_i p_i = 1\}$, in which $n$ is the number of classes $\mathcal{Y}=\{1, \ldots, n\}$ and $p_i$ is the prior probability (a.k.a.\ ``class frequency'', or ``class prevalence'') of class $i$ in  bag $\mathbf{X}$. Given a classifier $\phi : \mathcal{X} \rightarrow \mathcal{Y}$,
and a bag $\mathbf{X}$, CC is defined as
\begin{align}
 \operatorname{CC}(\mathbf{X})_{i}=\hat{p}_i=\frac{1}{|\mathbf{X}|} \sum_{\mathbf{x}\in\mathbf{X}} \ind[\phi(\mathbf{x})=i]
 \label{eq:cc}
\end{align}
%
%
\namedpar{A caveat on terminology}. This paper integrates concepts from different disciplines (quantification, information retrieval, and fairness), each of which employs its own consolidated terminology. 
Thorough this paper, we will interchangeably use the terms ``classes'' (here denoted by $\mathcal{Y}$) and ``groups'' (often denoted by $\mathcal{A}$ in the fairness literature), ``labels'' and ``sensitive attributes''. 
The reader should also note that, despite referring to different concepts, we might interchangeably use ``bags'' (or ``multi-sets'') and ``ranked lists of \replace{documents}{items}'', since our quantifiers  regard the latter as unordered objects.


\namedpar{Dataset shift}. The essence of quantification is that of tackling a situation in which there is a change (``shift'', or ``drift'') 
between the distribution $P_{tr}$ from which instances used to train the quantifier have been drawn and the distribution $P_{te}$ from which the test data are drawn. 
\replace{The reason is that in the absence of any such shift (i.e., if the IID assumption holds), predicting the test distribution becomes trivial, since the training prevalence is already a good estimate of the test prevalence.}{In online hiring, this corresponds to a realistic setting where the training set, consisting of candidates who disclose their sensitive data $y$, is not sampled IID from the same distribution as the test set.} The scenario in which $P_{tr}(X,Y)\neq P_{te}(X,Y)$ is generally known as \textit{dataset shift}~\cite{Storkey:2009lp}. 

Although $P_{tr}$ and $P_{te}$ are rather standard notation in machine learning for referring to the training and test distributions, throughout this paper we will consider more than two such distributions. For this reason, we will use the nomenclature $P_A$ to refer to the distribution from which an empirical sample of data items $A$ has been drawn. In this way, we use $P_L$ to denote the distribution from which \underline{l}abelled documents are drawn, and $P_U$ to denote the distribution from which the \underline{u}nlabelled documents are drawn. Further distributions will be introduced when needed.

Among the main types of shift that have been described in the literature, quantification has traditionally focused on \emph{prior probability shift} (PPS). This type of shift is characteristic of \textit{anti-causal learning}~\cite{Scholkopf:2012je} -- i.e., learning problems in which the covariates represent symptoms of the phenomenon we want to predict, and that are typically modelled via the factorization $P(X,Y)=P(X|Y)P(Y)$ -- and is characterized by the fact that $P_{L}(Y)\neq P_{U}(Y)$ while $P_{L}(X|Y)=P_{U}(X|Y)$. 

\namedpar{A simple quantifier}. Arguably, the simplest quantification method devised to counter PPS is the so-called \textit{Adjusted Classify \& Count} (ACC)~\cite{Forman:2005fk}. ACC is better described in the binary case $\mathcal{Y}=\{0,1\}$ (the multiclass extension is straightforward), by observing that
%
\begin{align}
\begin{split}
 P_{U}(\hat{Y}=1) & = P_{U}(\hat{Y}=1|Y=1)P_{U}(Y=1) \\ 
 & + P_{U}(\hat{Y}=1|Y=0)P_{U}(Y=0)
\end{split}
\label{eq:acc-obs}
\end{align}
\noindent where $P_{U}(\hat{Y}=1)$
corresponds to the CC estimate of the positive class and 
$P_{U}(\hat{Y}=1|Y=1)$ and $P_{U}(\hat{Y}=1|Y=0)$
are the \emph{true positive rate} (tpr) and \emph{false positive rate} (fpr) of the classifier 
\alexcomment{$\phi$.} 
These two quantities can be estimated using the training data given that the class-conditional distributions of the training and test data are assumed invariant. ACC is thus defined as
\begin{align}
 \operatorname{ACC}(\mathbf{X})_{1}=\frac{\operatorname{CC}(\mathbf{X})_{1}-\hat{\mathrm{fpr}}}{\hat{\mathrm{tpr}}-\hat{\mathrm{fpr}}}
 \label{eq:acc}
\end{align}
\add{In online hiring, ACC can estimate prevalence of different ethnicities in sets of candidates.}

\subsection{Countering Sample Selection Bias}
\label{sec:pitfall}
%
\noindent \remove{Given that many of the proposed evaluation measures for assessing fairness in rankings rely on estimating group distributions, one might reasonably expect that simply incorporating some of the most sophisticated quantification methods would enhance the accuracy of metric prediction. However, contrary to this expectation, our preliminary experiments have shown otherwise. Indeed, we observed that the application of quantification techniques almost always led to a deterioration in the ranking fairness prediction with respect to CC.}

\remove{The reason is that, while it makes sense to think that the prior distribution of the groups might change over time, or on a per-query basis, the truth is that the case we are facing here is not an example of PPS. To see why}

\namedpar{Origin}. Consider the random variable $Q$ that takes on values 1 (``the \replace{document}{item} is relevant'') and 0 (``the \replace{document}{item} is irrelevant'') with respect to a specific query. Note that the class-conditional distribution is a mixture of relevant and irrelevant \replace{documents}{items}, i.e., $P(X,Q|Y)=P(X|Y,Q=1)P(Q=1)+P(X|Y,Q=0)P(Q=0)$; however, if $U_q$ are the test documents retrieved for a query, we might expect 
$P_{L}(Q=1)\ll P_{U_q}(Q=1)$,
since it is likely that the vast majority of the training data used for learning our quantifier is irrelevant to a specific query, while the majority of the \replace{documents}{items} retrieved for the query are indeed relevant to it. 
The class-conditional distributions are thus different, and this clashes with the PPS assumptions.

Note that the random variable $Q$ might be regarded as the ``selection variable'', which is representative of a type of dataset shift known as \emph{sample selection bias} (SSB).\footnote{Sample selection bias is often defined differently, since the selection variable has an effect on the way the training instances (and not the test instances, as in our case) are selected~\cite{Storkey:2009lp}.} 
While different quantification methods have shown varying degrees of effectiveness in addressing different types of shift~\cite{gonzalez2024binary}, we are unaware of \remove{the existence of} any quantification method robust to SSB. 
In the next section, we propose the first approach to make quantification methods robust against SSB\add{, thereby adapting it for QFE}.



%
\remove{\noindent The reason why traditional quantifiers failed in the presence of SSB in our preliminary experiments has to do, as previously hinted, with the violation of the assumption $P_{tr}(X|Y) = P_{te}(X|Y)$, which characterizes PPS.}

\namedpar{Mitigation}. Let us turn back to the ACC method (Equation~\eqref{eq:acc}) to illustrate the problem (a similar rationale applies to other quantification algorithms as well, as we discuss in Section~\ref{sec:multiclass}). 
Recall that ACC replaces 
$P_{U}(\hat{Y}=1|Y)$ with $P_{L}(\hat{Y}=1|Y)$
in Equation~\eqref{eq:acc-obs} on the grounds that the class-conditional distributions of training and test datapoints are invariant. That 
$P_{L}(X|Y) = P_{U}(X|Y)$ implies $P_{L}(\hat{Y}|Y) = P_{U}(\hat{Y}|Y)$
follows from the fact that $\hat{Y}$ depends uniquely on $X$ by means of a function (the classifier); inasmuch as the classifier is a measurable function this equivalence holds~\cite{Lipton:2018fj}.

In general, 
$P_{U}(\hat{Y}=i|Y=j)$
represents the \emph{classification rates} of the classifier in the \emph{test set}, and is given by
\begin{align}
 P_{U}(\hat{Y}=i|Y=j) = \underset{\x\sim P_{U}(X|Y=j)}{\mathbb{E} \left[ { \mathds{1} \left[ \phi(\x)=i \right] } \right] }
\end{align}
%
\noindent Of course, we do not have access to the true distribution 
$P_{U}(X|Y=j)$
of the expectation, but if we could assume 
$P_{U}(X|Y)=P_{L}(X|Y)$,
then this expectation could be estimated by means of an empirical distribution 
$\x_1,\ldots,\x_m\sim P_{L}(X|Y=j)$,
as
\begin{align}
 P_{U}(\hat{Y}=i|Y=j) \approx \frac{1}{m}\sum_{k=1}^{m} \mathds{1} \left[ \phi(\x_k)=i \right] 
\end{align}
%
\noindent Although we know that this assumption is flawed in the presence of SSB (Section~\ref{sec:pitfall}), one fundamental observation arises: the pitfall stems from the choice of the empirical distribution used to characterize the classifier 
$\phi$,
rather than from the classifier itself.

This is important since most quantifiers use a training set $L=\{(\x_i,y_i)\}_{i=1}^m$, $\x_i\in\mathcal{X}$, $y_i\in\mathcal{Y}$ to both learn a classifier 
$\phi$,
\emph{and} estimate, via cross-validation,\footnote{This is in order to avoid the same datapoint being classified to take part in the training of the classifier.} its classification rates (in our binary example, this reduces to estimating tpr and fpr). 
Note also that training a classifier is costly, whereas learning the classification rates is rather inexpensive as it only involves issuing predictions and rearranging counts.

\namedpar{Main idea}. We propose to disentangle the classifier-training phase from the correction learned by the quantifier. We therefore assume to have access to two sets of labelled data, 
$L_{\phi}=\{(\x_i,y_i)\}_{i=1}^m$
that we use to train our classifier (offline since it is costly), and \replace{$L_{q}=\{(\x_i,y_i)\}_{i=1}^{m'}$}{$L_{corr}=\{(\x_i,y_i)\}_{i=1}^{m'}$} that we use to learn the correction (at query time since it is inexpensive). What remains to make quantification robust to SSB is to counter the selection bias between $L_{q}$ and $L_{corr}$.

\begin{figure}
    \centering
    \includegraphics[width=0.8\linewidth]{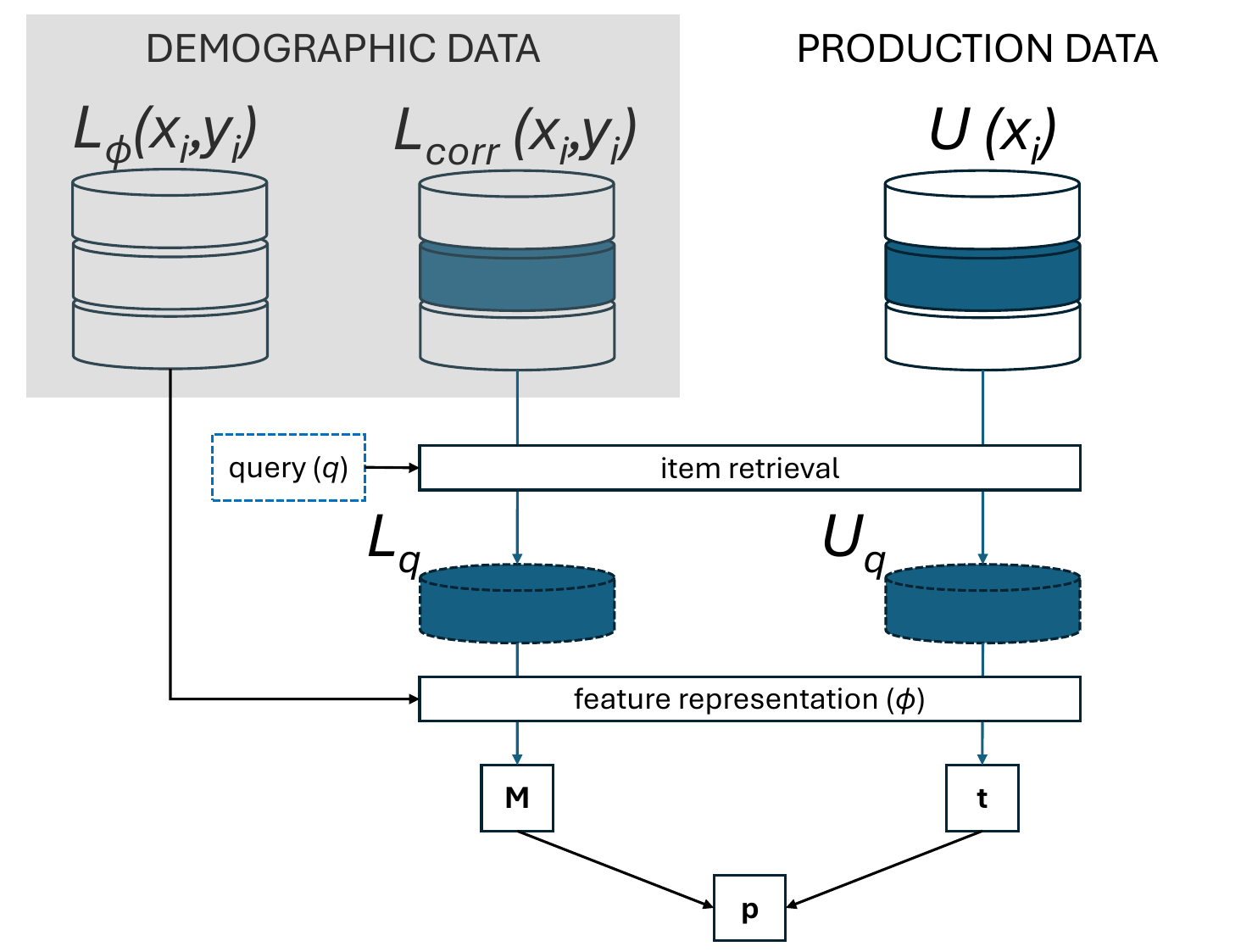}
    \caption{Schematic of our proposed approach for Query Fairness Evaluation. Demographic data is split into $L_{\phi}$, used to learn a feature representation function $\phi$, and $L_{corr}$, employed for query-specific corrections. For each query $q$, the prevalence $\mathbf{p}$ of sensitive attributes in the retrieved production data $U_q$ is estimated by representing ranked items according to $\phi$ and leveraging a subset $L_q$, containing the top-ranked items of $L_{corr}$, to learn a correction factor $\mathbf{M}$ specific to $q$.}
    \label{fig:diagram}
\end{figure}

Figure \ref{fig:diagram} summarizes our approach, with \textbf{M} and \textbf{t} denoting item representations explained Section \ref{sec:multiclass}. Reducing the sampling bias from the test data is impossible; the test \replace{documents}{items} are retrieved by the search engine precisely to guarantee that \replace{documents}{items} are relevant to a specific query. The main idea we \replace{explore}{propose} in this paper is \emph{to use the same search engine with the same query} to retrieve, from an \emph{auxiliary} pool of training documents \add{$L_{corr}$} (\replace{different from those used to train the classifier}{with 
$L_{corr} \cap L_{\phi} = \varnothing$
}), a subset of training items $L_q \subset L_{corr}$ that are biased towards the query similarly to items in $U_q$. This way, the empirical distribution $L_{q}$ that we retrieve from the auxiliary pool ($L_{corr}$) can now be regarded as a sample from a query-biased distribution 
$P_{L_q}$
and, since we can now assume 
$P_{L_q}(Q)\approx P_{U_q}(Q)$
(i.e., both distributions are biased towards the query) then we can also assume 
$P_{L_q}(X,Q|Y)\approx P_{U_q}(X,Q|Y)$
and restore the fundamental PPS assumption. Given that the SSB in $L_{q}$ mimics the SSB that affects the test data, \emph{the sampling bias shift vanishes}.

We thus consider an auxiliary set of labelled \replace{documents}{items} \replace{$\mathcal{L}$}{$L_{corr}$} containing pairs $(\x_i,y_i)$ labelled by sensitive attributes (hereafter called the 
``correction pool'').
From this \replace{correction pool ($\mathcal{L}$)}{set $L_{corr}$}, we select, using the same retrieval model and query that we issue on the test pool ($U$), a ranked list of (labelled) \replace{documents}{items} $L_q$ that we use to learn a per-query quantification correction to estimate group prevalence in the top-$k$ prefixes of (unlabelled) rankings $U_q$. \add{Considering ACC applied to online hiring as an example, to estimate the prevalence of different ethnicities in search results $U_q$ for a given query, we compute the classification rates for Equation \eqref{eq:acc} from a subset $L_q \subset L_{corr}$ of top-ranked items, instead of using the whole labelled set $L_{corr}$.}
\alexcomment{Do we use ``labeled'' or ``labelled''; I am encountering both here and there, and replacing them in favour of ``labelled''.} \afabcomment{ok}

\subsection{Quantifying Query Fairness}
\label{sec:multiclass}

\noindent In the previous section, we have explained the intuitions behind our method with respect to (binary) ACC, a relatively simple quantifier.
In this section, we generalize the rationale to more sophisticated multiclass quantification methods.

In the modern perspective of multiclass quantification~\cite{Bunse:2022ky}, most quantifiers can be framed as the problem of solving for $\mathbf{p}\in\Delta^{n-1}$ \add{(unknown prevalences)} the  system of linear equations
\begin{align}
 \mathbf{t}=\mathbf{M}\mathbf{p}
 \label{eq:framework}
\end{align}
\noindent where $\mathbf{t}=\Phi(U)$ is the representation of the test bag $U$, and 
$\mathbf{M}=[\Phi(L^1_{corr}),\ldots,\Phi(L^n_{corr})]$ is the matrix containing the class-wise representations of the class-specific 
correction sets $L^i_{corr}=\{\x_k : (\x_k,y_k)\in L_{corr}, y_k=i\}$ , for a given representation function $\Phi : \mathbb{N}^{\mathcal{X}} \rightarrow \mathbb{R}^z$ that embeds bags into $z$-dimensional vectors, for some $z$. 

Most quantifiers rely on a representation function of the form:
\begin{align}
 \Phi(\mathbf{X})=\frac{1}{|\mathbf{X}|}\sum_{\x\in\mathbf{X}} \phi(\x)
\end{align}
\noindent in which a surrogate instance-wise representation function $\phi : \mathcal{X} \rightarrow \mathbb{R}^z$ is invoked, thus effectively computing a mean embedding. Here, we deliberately use $\phi$ to denote both the representation function and a classifier, since most quantification methods define the representation function upon the output generated by a classifier.

Different choices for $\Phi$ and $\phi$ give rise to different instances of quantification methods. For example, ACC comes down to choosing, as our representation function $\phi$, the output of a crisp classifier 
encoded as a one-hot vector. The columns of $\mathbf{M}$ thus represent the classification rates of the classier (as estimated on correction data), and the problem comes down to reconstructing the class counts of the test examples as a linear combination of 
\afabcomment{the columns in $\mathbf{M}$}, by solving $\mathbf{p}=\mathbf{M}^{-1}\mathbf{t}$.

More sophisticated methods exists. For example, \textit{Probabilistic Adjusted Classify and Count} (PACC)~\cite{Bella:2010kx} (that we use in our experiments), defines $\phi$ as a probabilistic classifier returning the posterior probabilities for each class. When $\mathbf{M}$ is not invertible~\cite{Bunse:2022oj}, the variant we employ frames Equation~\eqref{eq:framework} as the minimization problem
\begin{align}
 \mathbf{p}'=\argmin_{\mathbf{p}\in\Delta^{n-1}} |\mathbf{t}-\mathbf{M}\mathbf{p}|^2
 \label{eq:pacc}
\end{align}
\noindent We also use KDEy~\cite{Moreo:2024KDEy}, a state-of-the-art multiclass quantification method that defines $\Phi$ as a Gaussian mixture model, obtained via kernel density estimation, of the posterior probabilities returned by a probabilistic classifier. We use the maximum-likelihood variant that solves Equation~\eqref{eq:framework} as the minimization problem
\begin{align}
 \mathbf{p}'=\argmin_{\mathbf{p}\in\Delta^{n-1}} \mathcal{D}_{\mathrm{KL}}(\mathbf{t}||\mathbf{M}\mathbf{p})
 \label{eq:kdey}
\end{align}
\noindent where $\mathcal{D}_{\mathrm{KL}}$ is the well-known Kullback-Leibler divergence.

Note that PACC and KDEy both rely on a probabilistic classifier that is trained in advance 
on $L_{\phi}$.
Equations~\eqref{eq:pacc} and \eqref{eq:kdey} only require converting the items in $L_{q}$ and the test items $U_q$ (both retrieved for the same query but from different pools, $L_{corr}$ and $U$, respectively) into posterior probabilities and solving the optimization problem. Both operations are rather fast given modern optimization routines \replace{and given the fact that the number of retrieved \replace{documents}{items} (either for training and test) is typically small (in our experiments, this number is bounded by 1000 items)}{(Section \ref{sec:times})}. 

\section{Experimental Setup}
\label{sec:experimentalsetup}
%
\noindent In this work, we aim to provide answers for the following research questions: 
\begin{itemize}
 \item \textbf{RQ1}: Does our new method improve over existing baselines? 
\item \textbf{RQ2}: Do quantification techniques allow for accurate QFE in a multiclass setting?
 \item \textbf{RQ3}: To what extent does the performance of our algorithm depend on the size of the correction pool and the rank exposure? 
\end{itemize}
The code that implements all our proposed and all baseline methods, and reproduces our experimental results, is publicly available.\footnote{\url{https://github.com/AlexMoreo/query-fairness-estimation}}


\subsection{Dataset \& Retrieval}
\label{sec:dataset}
\noindent To answer our research questions, we use the TREC 2022 Fair Ranking Track collection~\cite{trec-fair-ranking-2022}. This dataset was established for the TREC Fair Ranking Track, in which the fairness of rankings produced by IR systems across different queries and various groups of documents with multiple classes is evaluated. It includes 6.5M English-language Wikipedia articles labelled with group information for various sensitive attributes. We choose this dataset over alternatives from the hiring domain since it is publicly available, it is large, and it encodes several multi-valued sensitive attributes.

\replace{To investigate the generalisation of our approach, we select multiple attributes for our evaluation. From the available attributes, we select those where a standard classifier demonstrates reasonable accuracy (by ``reasonable'' we here consider above 0.75 accuracy). Table~\ref{tab:accucary_classifiers}  shows the performance over different attributes when linear regression (LR) and support vector machine (SVM) are used as classification models. The table indicates that, for three out of five attributes (``Geographic Location'', ``Gender'', and ``Age of Topic''), both classification models achieve high accuracy (>$0.75$). However, for the ``Popularity'' and ``Languages'' attributes, neither classification model surpasses the 0.75 threshold. In our experiments, we focus on the three attributes and their groups (top-3 rows in Table~\ref{tab:accucary_classifiers}) for which we have observed sufficient performance of the classifiers.}{To analyse the generalisation of our approach, we select multiple attributes for our evaluation, namely geographic location, gender, and age of topic. Standard classifiers such as logistic regression and SVM demonstrate reasonable accuracy ($>$ 0.75) for these attributes.}


%

In addition to the documents and their demographic labels, the collection includes 97 queries. \replace{To be able to issue the queries on the collection, w}{W}e index the documents using a Porter stemmer and stop word removal with the help of PyTerrier~\cite{macdonald2021pyterrier}. As our retrieval model, we use BM25~\cite{robertson1995okapi} with its standard parameters. \remove{We use these settings over all configurations of the correction pool and the test set.}  


\subsection{Evaluation measures}
\label{sec:measures}
\noindent In our experiments, we assess the accuracy of QFE on multiclass groups. We concentrate on metrics that implement an exposure drop-off based on rank position, thereby reflecting the bias by which users tend to pay more attention to documents ranked higher~\cite{craswell2008experimental}.
As our metric of query fairness we rely on the normalized discounted Kullback-Leibler divergence (rKL)~\cite{zehlike2022fairness} which, for a given set of retrieved 
documents $U_q$ and different ranking levels $k\in K$ (we consider $K=\{50,100,500,1000\}$), is given by 
\begin{align}
 \mathrm{rKL}(U_q)=&\frac{1}{Z} \sum_{k\in K} \frac{1}{\log_2 k} \mathcal{D}_{\mathrm{KL}}(\mathbf{p}^{k}||\mathbf{p}^{*})
 \label{eq:kld}
\end{align}
\noindent where $\mathbf{p}^{k}$ is the group distribution for the top-$k$ documents, and $\mathbf{p}^{*}$ is the group distribution in all judged-relevant documents in the test collection $U$ from which the ranked list $U_q$ is retrieved. $Z$ is simply the normalization factor computed as $Z=\sum_{k\in K} 1 / \log_2 k$. \remove{The KL-divergence between two discrete distributions $\mathbf{p},\mathbf{q}\in\Delta^{n-1}$ is given by:} 
%

In Section~\ref{sec:binary} we will also test our methods against other baseline methods that are binary-only and that are tailored to one specific fairness metric called normalized discounted difference (rND)~\cite{ghazimatin2022measuring} defined by
\begin{align}
 \mathrm{rND}(U_q)=\frac{1}{Z} \sum_{k\in K} \frac{1}{\log_2 k} \left| \; p^{k}_1 - p^{*}_1 
 \; \right|
\end{align}
\noindent Note that $\mathrm{rND}$ only considers the prevalence of the positive class, which is taken to be the prevalence of the protected or disadvantaged group.

Since we work under the unawareness assumption, $\mathbf{p}^{k}$ is unknown and needs to be estimated. We thus denote by $\hat{\mathrm{rKL}}(U_q)$ (resp. $\hat{\mathrm{rND}}(U_q)$) the score obtained using predicted distributions $\hat{\mathbf{p}}^k$ in place of the true ones.
In order to assess the accuracy on the prediction of the fairness metric $M$ (be it $\mathrm{rKL}$ or $\mathrm{rND}$), we report the \textit{absolute error} averaged across all ranked lists $U_q$ retrieved for all queries ($Queries$), which is defined as
\begin{align}
 \mathrm{AE}(Queries,M)=\frac{1}{|Queries|}\sum_{U_q\in Queries} \left| M(U_q) - \hat{M}(U_q) \right|
\end{align}
\noindent \replace{Given that our methods are based on quantification, we}{We} also report the relative absolute error (RAE), a quantification-specific measure that confronts a predicted distribution with the true distribution, and is defined as
%
%
\begin{align}
\mathrm{RAE}(\mathbf{p}, \hat{\mathbf{p}}) = \frac{1}{n} \sum_{i=1}^n \frac{|\hat{p}_i - p_i|}{p_i}
\label{eq:rae}
\end{align}
\noindent \replace{Arguably, RAE is preferred over other existing evaluation measures used for quantification assessment~\cite{Sebastiani:2020qf} in the context we are facing here. The reason is that RAE effectively considers the errors for each class in proportion to the magnitude of the true prevalence, which is expected to be low for disadvantaged groups and is thus likely to be neglected by other evaluation measures. In contrast to fairness-specific metrics, RAE does not implement any exposure drop-off. We therefore report values of RAE at different cutoffs $k$.}{We choose RAE as it caters to minority classes by highlighting estimation errors that are small in absolute terms but proportionally large~\cite{Sebastiani:2020qf}.}



\subsection{Experimental protocol}
\label{sec:protocol}
\begin{figure}
 \centering
 \includegraphics[width=0.8\linewidth]{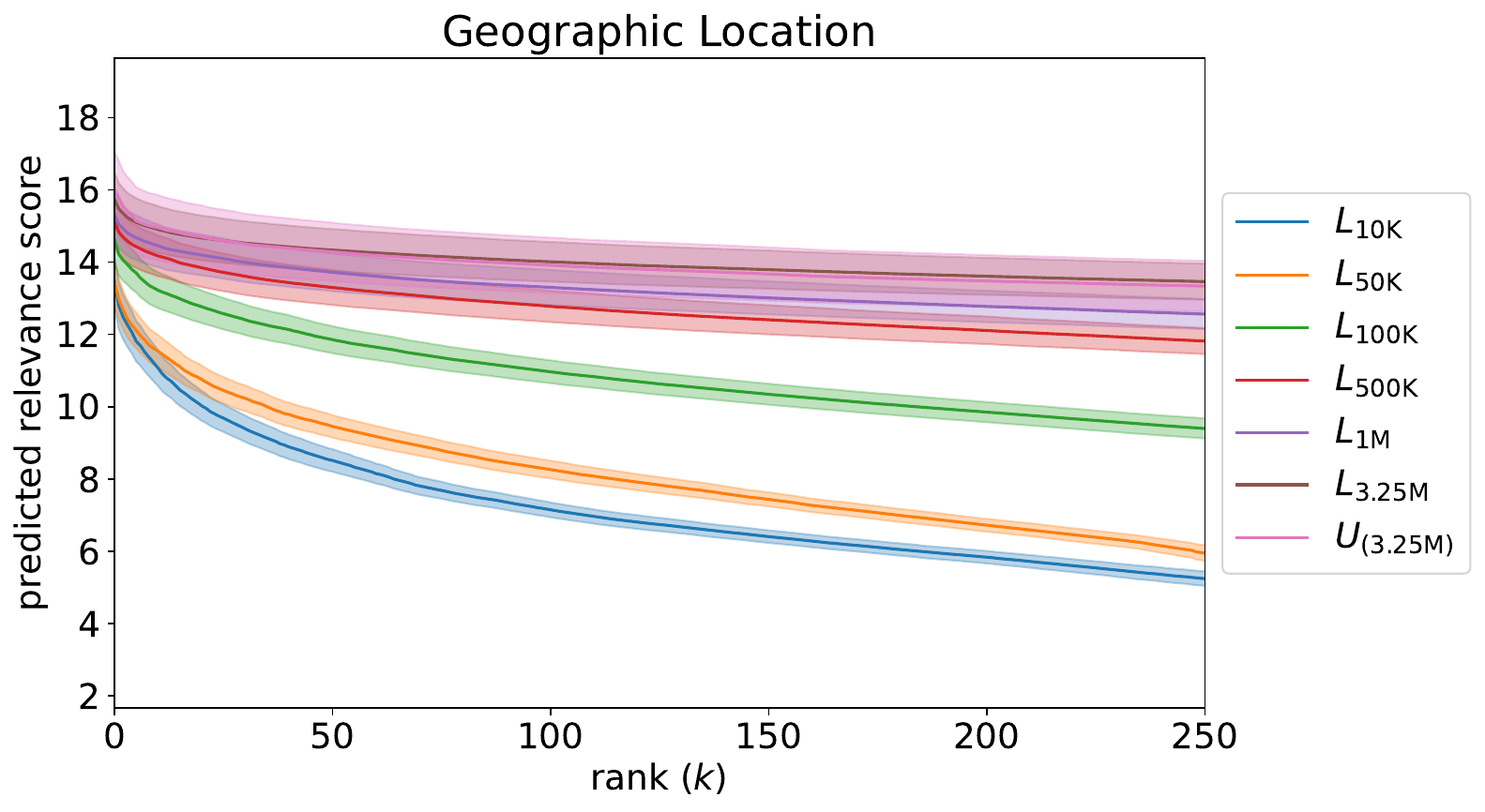}
 \caption{Distribution of predicted relevance score per document rank across all queries. \replace{Different curves represent different pools from which the documents are retrieved.}{As $|L_{corr}|$ increases its distribution aligns more closely with $U$.} 
 }
 \label{fig:rel_score_dist}
\end{figure}

\noindent In this section, we turn to describe the experimental protocol we have designed to provide answers to our RQs. The pseudocode describing our protocol can be consulted in Algorithm~\ref{alg:prot}.
\input{pseudocode}
The experimental variables we consider are listed below:
\begin{itemize}
 \item $size$: 
 The size of the correction pool $L_{corr}$ from which the documents $L_q$ are retrieved. This is important since there is an evident trade-off between cost and performance: a larger pool size implies higher labelling cost, while at the same time helps in reducing the discrepancy between the distribution of the documents retrieved from the training and test pool (Figure~\ref{fig:rel_score_dist}).\remove{; that is, it is more likely that the search engine manages to find relevant documents from a larger correction pool (see Figure~\ref{fig:rel_score_dist}). As a result, larger pool sizes correspond to more similar class-conditional distributions, which in turn should result in better quantification performance.} In Line~\ref{line:var_size} we let $size$ vary from a more realistic setting in which 10K labelled documents are assumed available, to the more optimistic (and unrealistic) scenario in which the correction pool is of the same size as the test pool (here corresponding to roughly 3.25M documents). We explore $size\in\{\mathrm{10K, 50K, 100K, 500K, 1M, 3.25M}\}$. \remove{The original pools ($\mathcal{L}, \mathcal{U}$) have a size of 3.25M each.
 The sets thus obtained are not independent of each other in the sense that the smaller pools are all subsets of the full correction pool (3.25M --Line~\ref{line:undersample}).}

 \item $k$: The rank cutoff. We investigate the impact the rank of the top-$k$ examined has in the accuracy of QFE. We let $k$ vary in the range $K=\{50,100,500,1000\}$ (Line~\ref{line:var_k}).

 \item $Queries$: We assess the performance of our methods in QFE across all 97 queries available in the TREC collection (Line~\ref{line:var_query}).

\end{itemize}



\subsection*{Protocol viewed from the quantification literature}

In quantification research, experimental evaluations often use a \textit{sampling generation protocol} to simulate shifts in class distributions, providing a stress test for assessing a quantifier’s performance. Typically, these shifts are applied to the test set. In our case, it is not clear how to achieve this without interfering with SSB. Still, we deem it important to impose such a shift in the prior since, in our case, the labelled and unlabelled pools are obtained by partitioning one preexisting collection. This results in the true underlying training and test distributions 
being nearly identical, 
which is an oversimplification of the problem. To avoid this oversimplification and test our quantifier under more realistic conditions, we introduce a shift by limiting the labelled data to a fixed number of documents per class (500 for 
$L_{\phi}$
in Line~\ref{line:hide_distribution_cls}, and 200 for $L_{q}$ in Line~\ref{line:hide_distribution_quant}).

\noindent \remove{In experimental evaluations carried out in quantification research, it is customary to artificially simulate varying amounts of shift, in order for stress-testing the ability that the quantifier will exhibit under deployment conditions when confronted with potentially widely different class distributions. 
Such variations are generally applied to the test set, by means of a so-called \emph{sampling generation protocol}.
In our case, it is not clear how to achieve a similar goal without interfering with SSB.
Still, we deem important to impose such shift in the prior, since the training and test pools we use are generated by partitioning one preexisting collection. The net effect is an oversimplification of the problem, in which the true distributions $P_{tr}$ and $P_{te}$ are actually the same.}

\remove{As a way out to solve this problem, we have decided to ``hide'' the true group distribution from the training side. To this aim, we draw a fixed number of documents (500) per class in order to train a classifier (Line~\ref{line:hide_distribution_cls}), and keep no more than a fixed number of documents (200) per class from the training documents retrieved for each query (Line~\ref{line:hide_distribution_quant}). 
In this way, we test the ability of our quantifiers to predict the true distribution while at the same time preventing the method from operating under overly mild conditions.}

\remove{In the future, it would be interesting to have access to differently characterized collections, in which the group prevalence might have varied naturally across training and test conditions. Having said that, this limitation in our protocol does not defy the fairness (in a broad sense) of our comparative results, since the same conditions affect all the tested methods and baselines in the same way.
}

\subsection{Methods}
\label{sec:baselines}
\noindent We experiment with our proposed variants:
\begin{itemize}

    \item PACC \cite{Bella:2010kx}: the ``Probabilistic Adjusted Classify \& Count'' quantification method described in Section~\ref{sec:multiclass}. Our proposed variant of PACC models the classification rates matrix of 
    $\phi$,
    using $L_q$.
    
    \item KDEy \cite{Moreo:2024KDEy}: the KDE-based quantification method described in Section~\ref{sec:multiclass}. Our proposed variant of KDEy models the class-wise densities of the posterior probabilities of $\phi$,
using $L_q$. 

\end{itemize}
\noindent We compare our methods to the following baselines:
\begin{itemize}


    \item Naive@$k$: a method that does not inspect the search results at all to estimate the prevalence of $U_q^k$ but simply reports the prevalence of $L_q^k$, i.e., of the top-$k$ training documents retrieved for each query from \replace{$\mathcal{L}_{size}$}{$L_{corr}$}. \remove{We include this ``naive'' baseline in our experiments to demonstrate the benefits that proper quantification methods exhibit when predicting the prevalence of $U_q^k$ from a correction learnt over $L_q$.}
    
    \item CC \cite{Forman:2005fk}: the ``Classify \& Count'' method described in Section~\ref{sec:method} and Equation~\eqref{eq:cc}. This method relies on the predictions issued by $\phi$, 
    and does not learn any correction based on $L_q$.
    
    

    \item PMC$_b$ \cite{ghazimatin2022measuring}: a binary-only method that corrects a preliminary estimate $\hat{\mathrm{rND}}_{\phi}$ obtained using ``proxy labels'' (i.e., labels predicted by $\phi$),
    by applying equation:
    \begin{equation}
        \mathrm{rND}(U_q) = \frac{\hat{\mathrm{rND}}_{\phi}}{(1-p)-w}
    \end{equation}
    \noindent 
    where 
    $p=P_{L}(\hat{Y}=1|Y=0)$ and $w=P_{L}(\hat{Y}=0|Y=1)$ 
    \cite<see Figure 1 (b) in>{ghazimatin2022measuring}. 

    \item PMC$_b^+$: a variant we propose for PMC$_b$ in which the correction factors $p$ and $w$ are not modelled on the same dataset $L$ used to train the classifier (as proposed by the inventors of the method), but from the items $L_q$ retrieved for each query. \remove{I.e., a variant that exploits our idea for modelling the correction.}

    \item PMC$_d$ \cite{ghazimatin2022measuring}: a binary-only method that corrects a preliminary estimate $\hat{\mathrm{rND}}_h$ 
    by applying equation:
    \begin{equation}
        \mathrm{rND}(U_q) = \hat{\mathrm{rND}}_h \cdot \left(\frac{(1-w)\cdot\beta}{x}-\frac{w\cdot\beta}{y}\right)
    \end{equation}
    where $p$ and $w$ are defined as for PMC$_b$, 
    $\beta=P_{L}(Y=1)$,
    and $x=(1-w)\cdot\beta+p\cdot(1-\beta)$ and $y=w\cdot\beta+(1-p)\cdot(1-b)$\remove{, under ``model-d'' assumption of the graphical model} \cite<see Figure 1 (d) in>{ghazimatin2022measuring}.

    \item PMC$_d^+$: a variant we propose for PMC$_d$ in which $p$, $w$, $\beta$, $x$, and $y$ are not modelled on 
    $L$ but on $L_q$.

\end{itemize}
%
%
%
\noindent Our implementations of quantification methods CC, PACC, and KDEy rely on the implementations available in QuaPy \cite{moreo2021quapy}, while the implementations of Naive@$k$, PMC$_b$, PMC$_b^+$, PMC$_d$, and PMC$_d^+$ are our own.

\namedpar{Classifier training}. 
All the methods we consider in this work, with the exception of Naive@$k$, rely on the outputs of a classifier. For the sake of a fair comparison, we use the same classifier in all cases. Our classifier of choice is Logistic Regression (LR), which is arguably becoming a \emph{de facto} choice in the field of quantification, due to the fact that many methods require probabilistic decisions and LR is known to deliver reasonably well-calibrated posterior probabilities 
\cite{schumacher2023comparative,moreo2021quapy}.
LR is trained offline, once and for all, for each category, on $L_{\phi}$. 

\namedpar{Model selection}.
We perform model selection on the hyperparameters of LR via 5-FCV; we explore the regularization strength $C\in\{10^i\}$ for $-4\leq i \leq 4$, and the \textsc{class\_weight} parameter in $\{Balanced, None\}$. 
The only model that has additional hyperparameters is KDEy, which depends on the ``bandwidth'' of the kernel. Of course, carrying out a model selection phase at query time is not viable. In order to select the bandwidth of KDEy we use 100 queries of the TREC 2021 Fair Ranking Track~\cite{trec-fair-ranking-2021} collection, a previous version of the TREC 2022 collection. We issue the queries on  
a 100K-sized correction pool to collect the training documents  for all attributes and classes, as well as on \replace{$\mathcal{U}$}{$U$} to collect our test rankings $U_q$.  We explored the bandwidth in the range $\{0.01, 0.02, \ldots, 0.10\}$.


\section{Results}
\label{sec:results}
\noindent In this section, we report and discuss the experimental results we have obtained.
Section~\ref{sec:binary} presents a comparison  against the PMC variants (discussed in Section~\ref{sec:related}); this experimental comparison is discussed separately because the above-mentioned models are binary-only (RQ1).
In Section~\ref{sec:multiclassresults} we show our main set of experiments, in which we assess the effectiveness of QFE considering multiclass groups (RQ2). Section~\ref{sec:variationsofk} analyses the extent to which the performance of our proposed methods depends on the rank $k$ and the correction pool size (RQ3).
Finally, Section~\ref{sec:times} reports averaged time measurements of our methods.


\subsection{\replace{Comparing against previous art: the binary case}{Binary protected attributes}}
\label{sec:binary}
\noindent 
In this section, we compare our proposed approach to previous related work (RQ1). 
\replace{As in the previous section, the}{The} results we discuss in the next paragraphs correspond to the more realistic scenario in which $size=10$K.
\remove{To the best of our knowledge, the only methods from the related literature that have addressed the problem of ranking fairness estimation under unawareness are those proposed by} \add{{We compare our proposed methods (PACC and KDEy) against the PMC variants}} \cite{ghazimatin2022measuring} Section~\ref{sec:baselines}. 


\replace{As recalled from the related work section, the PMC variants, like our methods, include a correction phase and rely on underlying classifiers. However, our approaches apply a pre-correction to class prevalence estimates for computing a fairness metric, whereas the}{The} PMC methods apply a post-correction to the fairness metric score, making them \replace{specific to one metric (rND)}{metric-specific} and binary-only.
In order to allow for an experimental comparison against the PMC variants, we produce binary versions of our datasets. In the binary setting, the positive class ($Y=1$) traditionally represents the minority or disadvantaged group. We thus binarize our datasets towards the following groups: ``Africa'' for Geographic Location, ``Female'' for Gender, and ``Pre-1900s'' for Age of Topic.
%
%
\noindent the rest of the groups are merged into the negative class ($Y=0$).

Table~\ref{tab:rND} reports the absolute error in the estimation of rND. The displayed values are averaged scores of the absolute error on the prediction of rKL (lower is better) across all 97 queries. Boldface indicates the best method for a given category. Superscripts $\dag$ and $\ddag$ indicate the methods (if any) whose scores are \emph{not} statistically significantly different from the best one at different confidence levels: symbol $\dag$ indicates $0.001 < p < 0.01$, while symbol $\ddag$ indicates $p\geq 0.01$ \afabcomment{Please double check.}\alexcomment{It is correct. I agree that it is confusing though; in other papers we are simply reporting one dag. Do you prefer to simplify it with only one dag?}\afabcomment{I think significance at 0.01 would be sufficient. Unless also reporting 0.001 adds something to the main message}. As the test for statistical significance, we rely on the non-parametric Wilcoxon signed-rank test. We use colour coding to facilitate the interpretation of the results, with green indicating the best result and red indicating the worst one per category.
\begin{table*}[tb]
 \centering 
 \caption{Results of QFE in terms of AE (lower is better) of rND prediction for \replace{$\mathcal{L}_{10\mathrm{K}}$}{$L_{10\mathrm{K}}$}
  (binary case). \add{KDEy beats all methods.}}
 \label{tab:rND}
 \resizebox{\textwidth}{!}{%
 \input{tables/rND_table.tex}
 }%
\end{table*}
From the analysis of the results the following observations can be drawn:
\begin{itemize}
    \item KDEy is the best-performing approach for the three categories. \remove{as well as on average.} 
    
    \item Although CC performs consistently worse than KDEy, the statistical test reveals these differences are not significant. \replace{That the errors produced by CC are small is an indication that the classifier is able to produce accurate predictions, thus leaving}{This may be due to high classifier accuracy in the binary case, leaving} small room for improvement for the correction phase of other methods. \add{Indeed CC fares significantly worse than KDEy in the multi-class settings (Section~\ref{sec:multiclassresults})}.
    
    \item PMC models perform worse than KDEy in a statistically significant sense in most cases. \remove{ and on average.} This is an indication that the assumptions upon which PMC models are built do not apply in QFE. 

    \item The variants PMC$^+$ we propose fare consistently worse than the original methods. 
    Intuitively, these variants should perform better, since the documents on which the correction is modelled are more similar to 
    the test data for which the correction is required.
    \replace{As stated previously, t}{T}his may be an indication that the post correction implemented by the PMC variants do not align with the characteristics of the distributions under consideration in QFE.

    \item Naive@$k$ performs badly in two out of three cases. This speaks in favour of the ability of KDEy to correct the class prevalence values, since the class prevalence of the top-$k$ training documents is not a good estimate for the top-$k$ test documents \emph{per se}.

    \item PACC falls short in terms of performance. A \replace{likely}{plausible} reason for this failure is the relatively low number of training documents used to model the classification rates\add{ (more on this in Section~\ref{sec:variationsofk})}. 
\end{itemize}



\subsection{\replace{Main experiments: the multiclass case}{Multiclass protected attributes}}
\label{sec:multiclassresults}
\noindent We now turn to query fairness estimation for multiclass sensitive attributes. Table~\ref{tab:rKL} reports the absolute error of rKL estimates (Equation \ref{eq:kld}) for a realistic scenario where the dataset size is set to $size=10K$.  Notational conventions are as in Table~\ref{tab:rND}. \add{Since ours is the first multiclass method, we compare our estimator against quantification baselines.}
The following observations emerge from our results:
\begin{itemize}
    \item KDEy consistently achieves the best performance compared to all other baselines, with statistically significant differences in the majority of cases \remove{on average} across the evaluated attributes. Furthermore, KDEy also exhibits the smallest standard deviation, indicating consistent performance.
    
    \item 
    Naive@$k$ performs erratically. In 
    Geographic Location, it achieves results similar to the best-performing method, KDEy. However, in the Gender category, it produces significantly higher errors compared to the best performer. This inconsistent performance is reflected in a high standard deviation across all categories.\remove{ Moreover, on average, Naive@$k$ yields the highest errors across all categories.}
    
    \item CC performs consistently worse than KDEy in all cases. \remove{and on average.} \replace{In two out of three cases, the differences in performance are not statistically significant, though, but only at a low confidence value.}{The differences are statistically significant at $p=0.01$; in one out of three cases they are significant at $p=0.001$.} \afabcomment{Please double check}
   
    \item PACC is, as in the binary case, not competitive with KDEy.
\end{itemize}

\begin{table}[b]
 \centering 
  \caption{Results of QFE in terms of AE (lower is better) of rKL prediction for
  \replace{$\mathcal{L}_{10\mathrm{K}}$}{$L_{10\mathrm{K}}$} (multiclass case). \add{KDEy beats all methods.}}
 \label{tab:rKL}
 \input{tables/rKL_table.tex}
\end{table}

The disparate outcomes we have obtained for PACC and KDEy deserve further analysis. Both methods rely on the same principle of deferring the correction-training phase at query time. As we will see in Section~\ref{sec:variationsofk}, though, PACC still performs decently in terms of quantification performance. 
Concerning RQ2 and in light of our observations, we can conclude that quantification techniques are indeed suitable for accurate QFE in a multiclass scenarios. \remove{In the next section, we compare our approach to existing baselines.}


\subsection{Variations of $k$ and size}
\label{sec:variationsofk}
\begin{figure*}[t]
 \centering
 \begin{subfigure}{0.295\linewidth}
 \includegraphics[width=\linewidth]{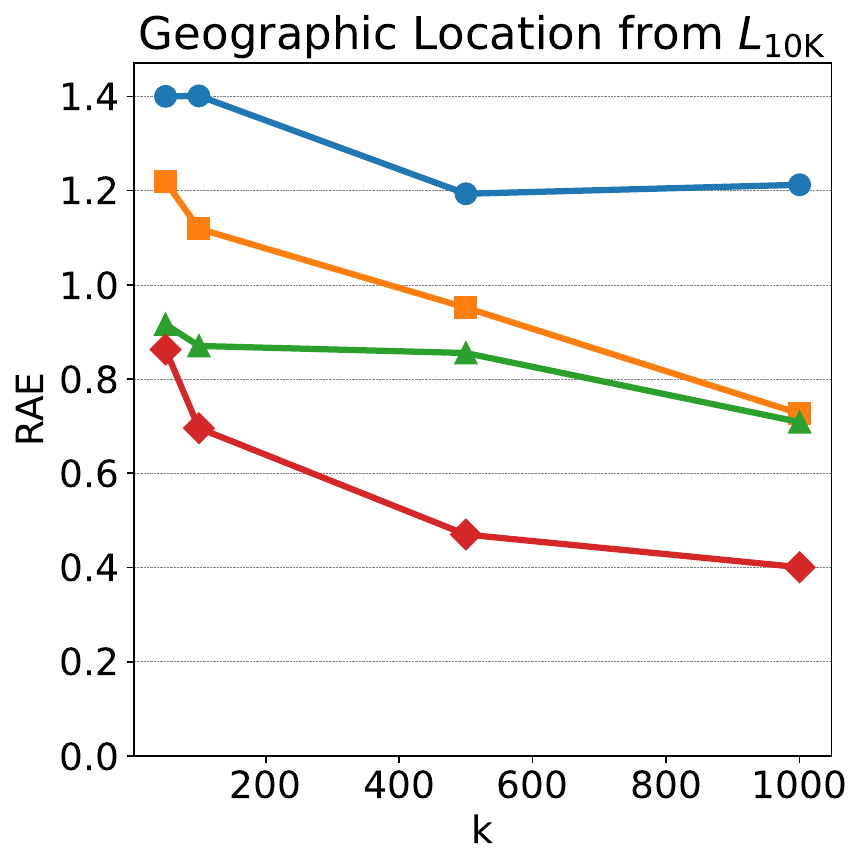}
 \end{subfigure}
 \begin{subfigure}{0.295\linewidth}
 \includegraphics[width=\linewidth]{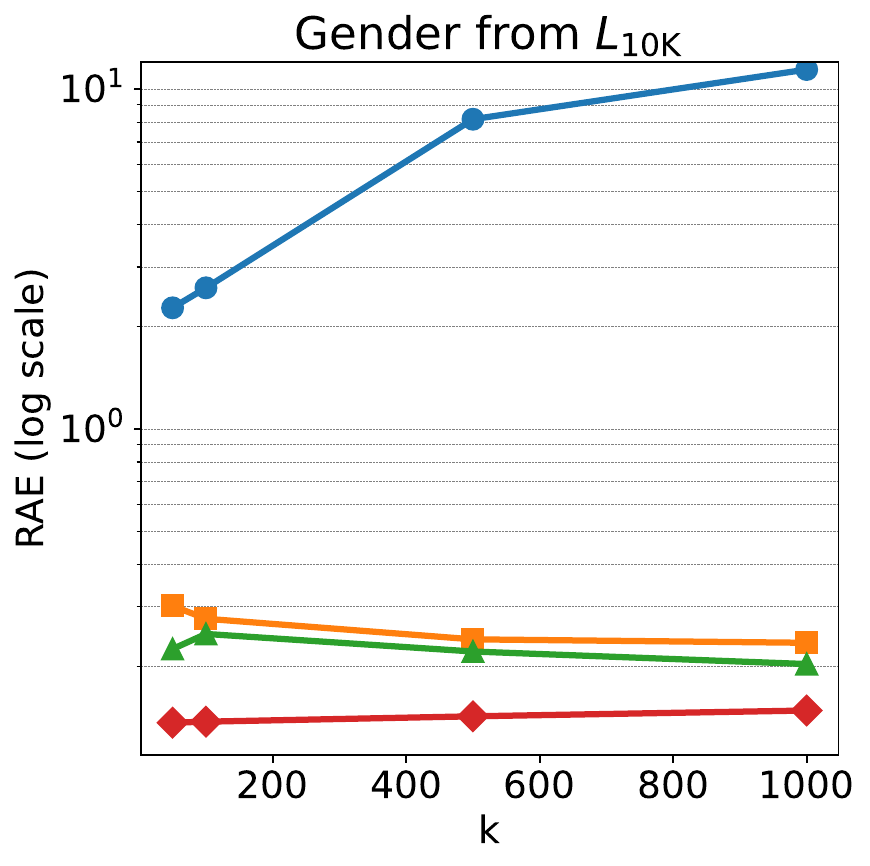}
 \end{subfigure}
 \begin{subfigure}{0.385\linewidth}
 \includegraphics[width=\linewidth]{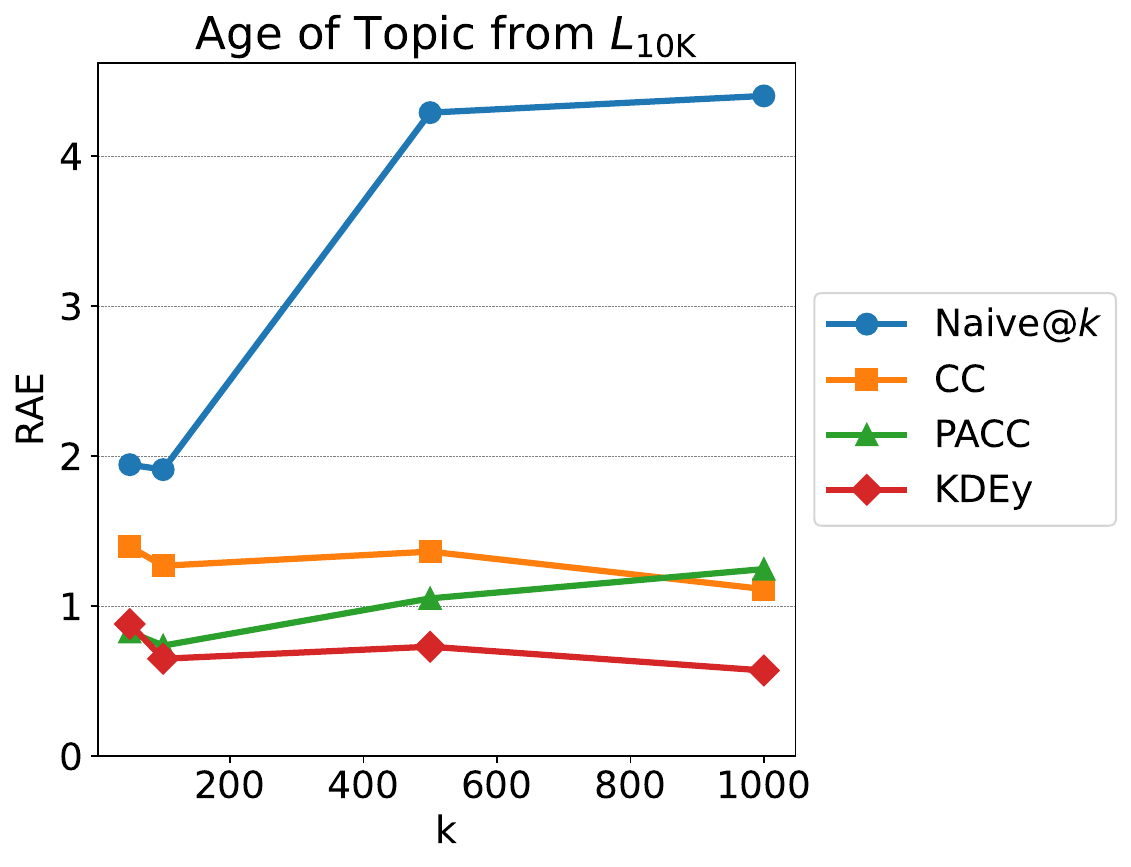}
 \end{subfigure}
 
 \caption{Variations in quantification performance (measured in terms of RAE -- lower is better) at different values of $k$. Note the log-scale in Gender. \add{KDEy and PACC outperform all baselines.} 
 }
 \label{fig:var_k}
\end{figure*}

\begin{figure*}[t]
 
 \begin{subfigure}{0.305\linewidth}
 \includegraphics[width=\linewidth]{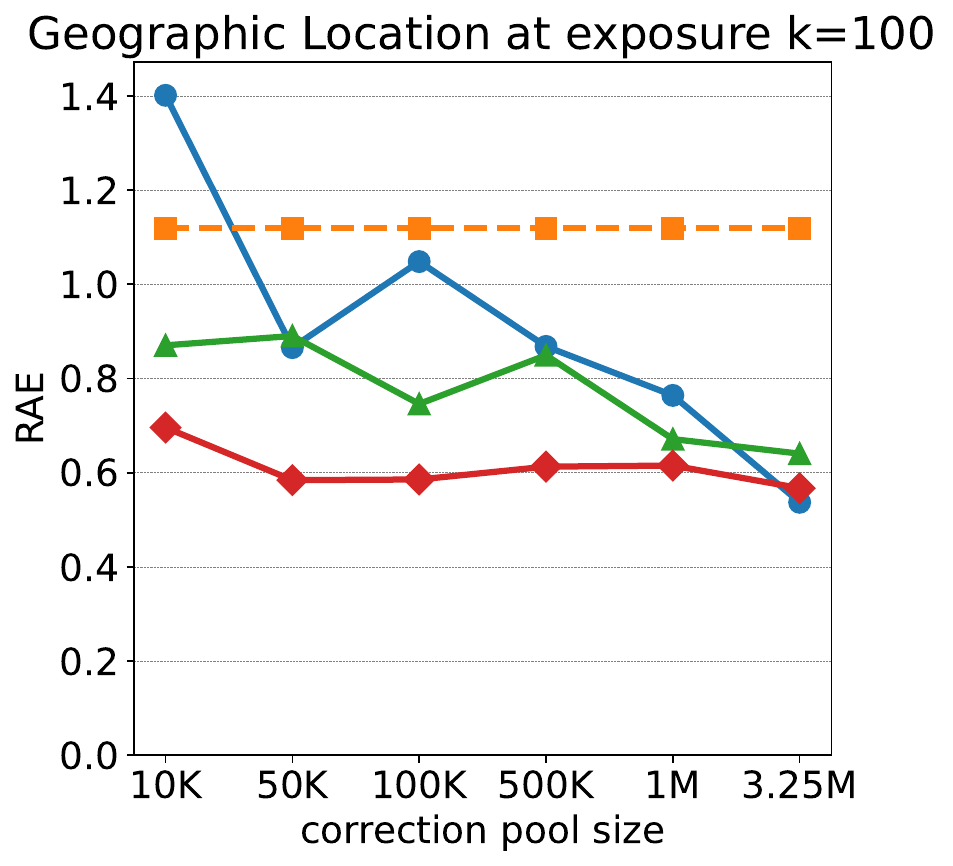}
 \end{subfigure}
 \begin{subfigure}{0.29\linewidth}
 \includegraphics[width=\linewidth]{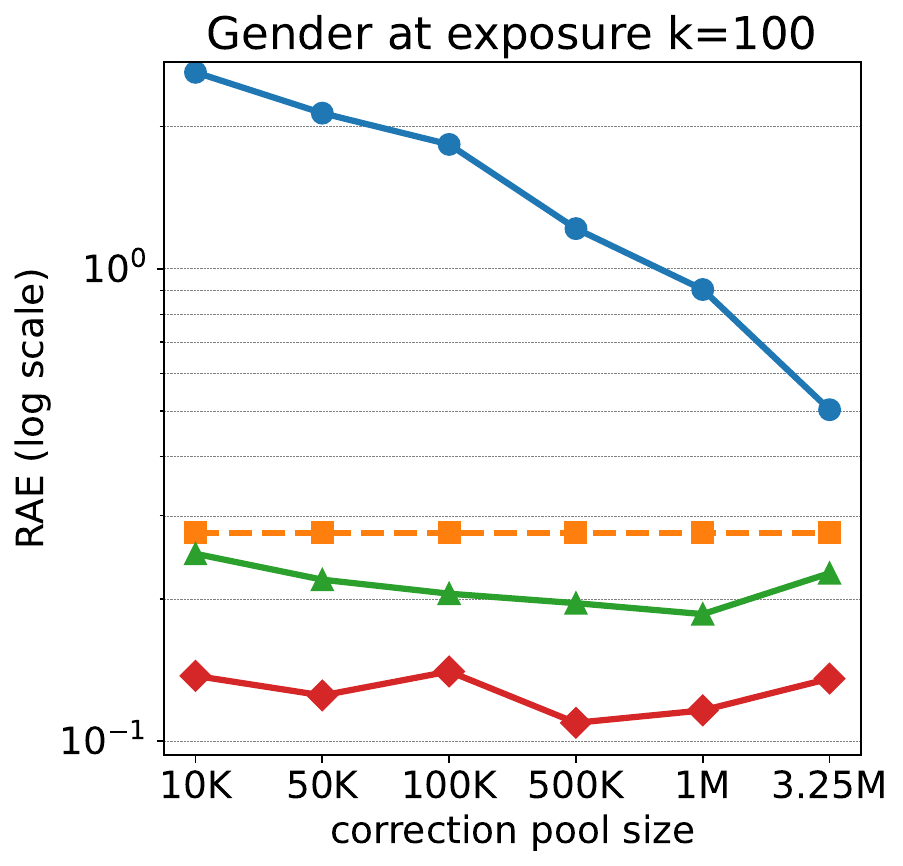}
 \end{subfigure}
 \begin{subfigure}{0.385\linewidth}
 \includegraphics[width=\linewidth]{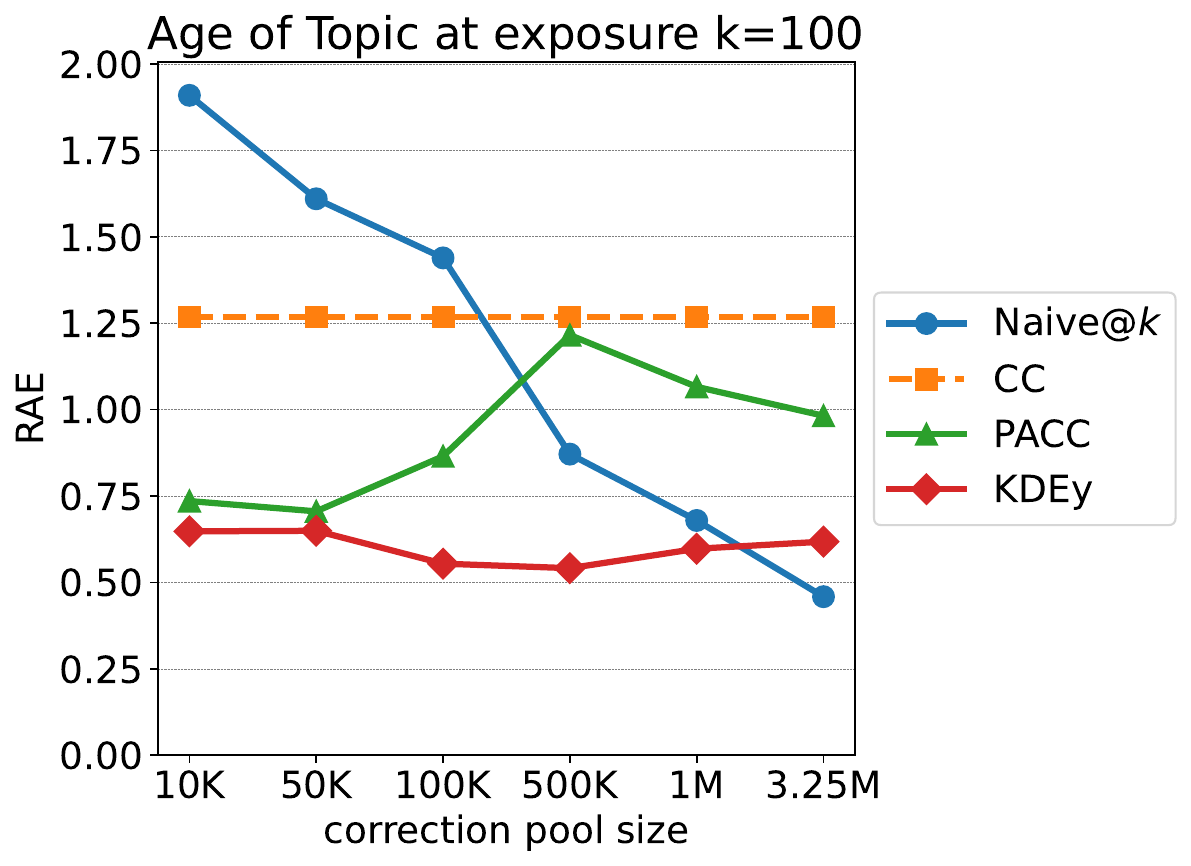}
 \end{subfigure} 

 \caption{Variations in quantification performance (measured in terms of RAE -- lower is better) at different correction pool sizes. Note the log-scale in Gender. \add{KDEy and PACC are more stable than Naive@$k$ and dominate CC.}
 }
 \label{fig:var_size}
\end{figure*}

\noindent In this experiment, we analyse the variations in quantification performance at different exposure levels $k$ and variations in the correction pool size (RQ3). 
We evaluate these experiments in terms of RAE (a quantification-specific measure) between the true distribution and the predicted distribution.
Figure~\ref{fig:var_k} displays variations in performance at different rank levels ($k\in\{50,100,500,1000\}$) for the case $L_{corr}\coloneq L_{10\mathrm{K}}$, while Figure~\ref{fig:var_size} displays variations in performance due to variations in the correction pool size at rank $k=100$.

%
These plots reveal some interesting findings.
First, PACC performs consistently better than CC in terms of quantification. 
This comes as a surprise, since the experiments reported in Table~\ref{tab:rND} and Table~\ref{tab:rKL} seemed to indicate PACC performs worse than \replace{the baselines}{CC}.

%
The key difference with respect to our previous experiments is the evaluation measure under consideration.
We argue that RAE is an appropriate measure in multiclass scenarios like the one we are facing here, given its ability to reflect the importance of an error with respect to the proportion of the true prevalence. This can be especially important in cases where one of the classes (the disadvantaged group) tends to display very low prevalence with respect to other classes (especially the privileged group). 

However, other evaluation metrics for QFE (e.g., rKL and rND) do not seem to align with the intuitions behind RAE.\footnote{Note that the KL-divergence in rKL also considers the ratio with respect to the target distribution. However, note that it also scales this factor by the predicted prevalence, which might simply be very close to 0, thus cancelling out the term in the aggregation; see Equation~\eqref{eq:kld}.}
In the future, it \replace{might}{will} be interesting to analyse the adequacy of a normalized-discounted-variant of RAE for fairness evaluation. We leave these considerations to future investigations.

Figure~\ref{fig:var_k} shows KDEy performs better than all other methods. Almost all methods show a tendency to improve as the exposure level increases. This is expected, 
as the quality of an estimated descriptive statistic (in this case: the prevalence) is known to depend on the size of the population under investigation. The only exception to this trend is Naive@$k$. The reason is that we actively tried to hide the original distribution (not only for Naive@$k$, but for all methods) by keeping no more than 200 documents per group in $L_q$ \add{(Line~\ref{line:hide_distribution_quant} in Algorithm~\ref{alg:prot})}. Moreover, Naive@$k$ and CC performing consistently worse than PACC and KDEy proves that our quantifiers are effectively learning a correction for the class counts of the classifier which is not spuriously based on the class distribution of $L_q$.

Figure~\ref{fig:var_size} also shows that Naive@$k$ has a clear dependency on the distributional similarity between \replace{$\mathcal{L}$ and $\mathcal{U}$}{$L_{corr}$ and $U$}, as witnessed by its drastic improvement when adding labelled data to the correction pool \add{$L_{corr}$, which makes it converge towards the same distribution as $U$}.
Conversely, the quantification performance of PACC and KDEy is relatively stable, and does not improve markedly when the amount of labelled data available in the pool increases. This indicates that our quantification-based methods are robust to drifts between \replace{$\mathcal{L}$ and $\mathcal{U}$}{$L_{corr}$ and $U$}. Moreover, this implies that the amount of labelling effort required to achieve reliable QFE scores is kept under reasonable bounds.
Note also that the CC method is represented as a flat curve. The reason is that CC does not leverage the data available in \replace{$\mathcal{L}$}{$L_{corr}$} by any means.

\subsection{Time Measurements}
\label{sec:times}

We measured the training and testing times of our Python implementations on a desktop computer equipped with a 12th Gen Intel(R) i9-12900K processor and 64GB of RAM, running Ubuntu 22. Our methods consist of a per-query correction learning phase followed by a phase of prevalence predictions. On average, PACC required 0.907 ms for learning and 3.316 ms for every prediction, while KDEy took 1.586 ms for learning and 6.395 ms for every prediction. Despite the requirement of a learning phase at query time, our methods demonstrate relatively fast performance and can be  integrated into the standard \replace{IR}{IAS} pipeline.


\section{Discussion and Conclusion}
\label{sec:conclusions}
In this work, we have investigated how to reliably assess the fairness of search results when demographic labels for ranked items are unavailable. We have demonstrated that simply counting over the predictions of a classifier leads to unreliable fairness assessments. To address this limitation, we have proposed \replace{the use of quantification}{novel quantification-based methods} to accurately estimate the prevalence of different groups in a ranking. 
The experimental evaluation \remove{on multiclass fairness groups of a publicly available fairness ranking benchmark }has shown that our approach can successfully predict query fairness, \add{including in the previously unaddressed multiclass case,} and \replace{do}{it does} so more accurately than existing methods \replace{from the literature}{in the binary case}.
While most quantification techniques are designed to counter prior probability shift, the problem at hand is instead mainly affected by sample selection bias. To the best of our knowledge, our approach is the first attempt towards making quantification robust to this type of shift naturally occurring in QFE.

\add{\namedpar{Limitations}. QFE may be challenging to integrate into learning-to-rank pipelines. First, leveraging protected attributes correction pools comes with infrastructural challenges of data integration. Second, conveying uncertainty of fairness estimates is an important problem we did not address in this work. Third, correction pools may exhibit particular properties (e.g. self-selection effects) that we did not explicitly model.}

\add{\namedpar{Future work}.} In future work, we aim to investigate the suitability of a normalised-discounted variant of RAE for fairness evaluation. We are also interested in exploring different collections where the group prevalence 
may have naturally varied across training and test conditions. \add{Finally, we aim to endow our QFE solutions with the ability to provide confidence intervals for the estimated values.}

\section*{Acknowledgments}

The work of AE has been supported by the SoBigData.it project (grant IR0000013). The work of FS has been supported by the FAIR project (grant PE00000013). The work of AF has been supported by the Alexander von Humboldt Foundation. The work of AM has been supported by the QuaDaSh project ``\textit{Finanziato dall’Unione europea - Next Generation EU, Missione 4 Componente 2 CUP B53D23026250001}''. The work of GM and IO has been supported in part by the Engineering and Physical Sciences Research Council grant number EP/Y009800/1, through funding from Responsible Ai UK (KP0011). 
\alexcomment{Missing projects for other authors?}

\end{document}

%% file: pseudocode.tex
\begin{algorithm}[t]
\LinesNumbered 
\SetNoFillComment
\footnotesize

    \SetKwInOut{Input}{Input} 
    \SetKwInOut{Output}{Output} 
   
    \Input{
        \textbullet\ \replace{TREC 2022 Fair Ranking Track collection}{data}  \\
        \textbullet\ Classifier learner CLS \\
        \textbullet\ Quantification method QUANT \\
        \textbullet\ Group labels \\
    } 
    \Output{
        \textbullet\ rKL of the fairness estimates \\
        \textbullet\ RAE of the group prevalence estimates  
    }

    \BlankLine

    $L, U \leftarrow$ split(\replace{TREC}{data}; 50\%, 50\%)

    $L_{\phi} \leftarrow$ draw($L$; 500 documents per group) \label{line:hide_distribution_cls}

    $L \leftarrow L - L_{\phi}$

    $\phi \leftarrow \mathrm{CLS}(L_{\phi})$ \label{line:classifier_training} 

    \For{$size \in \{ \mathrm{10K, 50K, 100K, 500K, 1M, 3.25M} \}$}{ \label{line:var_size}

        $L_{size}  \leftarrow$ undersample($L$; $size$) \label{line:undersample}

        $L_{corr} \coloneq L_{size} $ \quad \# alias

        \For{$\mathrm{query} \in$ Queries}{ \label{line:var_query}

            $U_q \leftarrow \mathrm{Retrieve}(U, \mathrm{query})@1000$ 

            $L_q \leftarrow \mathrm{Retrieve}(L_{corr}, \mathrm{query})@1000$ 

            $L_q \leftarrow $ keep($L_q$; max 200 most relevant documents per group) \label{line:hide_distribution_quant}

            $\lambda_h \leftarrow$ QUANT($L_q$, $h$)

            \For{$k\in\{50, 100, 500, 1000\}$}{ \label{line:var_k}

                $\hat{\mathbf{p}}^k \leftarrow \lambda_h(U_q^k)$

                $\mathbf{p}^k \leftarrow \mathrm{prevalence}(U_q^k)$

                compute $\mathrm{RAE}(\mathbf{p}^k, \hat{\mathbf{p}}^k)$
            
            }
        }

        compute $\mathrm{AE}(Queries, \mathrm{rKL})$

    }
\caption{Experimental protocol.}
\label{alg:prot}
\end{algorithm}

%% file: tables/rND_table.tex
\begin{tabular}{ccccccccc}
\toprule
\multicolumn{1}{c}{} & Naive@$k$ & CC & \add{PACC \footnotesize{(ours)}} & \add{KDEy \footnotesize{(ours)}} & PMC$_b$ & PMC$_b^+$ & PMC$_d$ & PMC$_d^+$ \\ \midrule
Geographic Location & .014$^{\dag}\pm^{\phantom{\dag}}$.023 \cellcolor{green!35} & .013$^{\ddag}\pm^{\phantom{\dag}}$.029 \cellcolor{green!38} & .043$^{\ddag}\pm^{\phantom{\dag}}$.143 \cellcolor{red!40} & \textbf{.012$^{\phantom{\dag}}\pm^{\phantom{\dag}}$.027} \cellcolor{green!40} & .020$^{\dag}\pm^{\phantom{\dag}}$.049 \cellcolor{green!20} & .030$^{\phantom{\dag}}\pm^{\phantom{\dag}}$.087 \cellcolor{red!6} & .020$^{\dag}\pm^{\phantom{\dag}}$.049 \cellcolor{green!20} & .031$^{\phantom{\dag}}\pm^{\phantom{\dag}}$.083 \cellcolor{red!7} \\
Gender & .047$^{\phantom{\dag}}\pm^{\phantom{\dag}}$.048 \cellcolor{red!40} & .014$^{\ddag}\pm^{\phantom{\dag}}$.042 \cellcolor{green!34} & .016$^{\ddag}\pm^{\phantom{\dag}}$.040 \cellcolor{green!29} & \textbf{.011$^{\phantom{\dag}}\pm^{\phantom{\dag}}$.026} \cellcolor{green!40} & .014$^{\ddag}\pm^{\phantom{\dag}}$.041 \cellcolor{green!33} & .015$^{\ddag}\pm^{\phantom{\dag}}$.037 \cellcolor{green!31} & .014$^{\ddag}\pm^{\phantom{\dag}}$.041 \cellcolor{green!33} & .015$^{\ddag}\pm^{\phantom{\dag}}$.037 \cellcolor{green!30} \\
Age of Topic & .040$^{\phantom{\dag}}\pm^{\phantom{\dag}}$.040 \cellcolor{red!26} & .025$^{\ddag}\pm^{\phantom{\dag}}$.040 \cellcolor{green!16} & .019$^{\dag}\pm^{\phantom{\dag}}$.023 \cellcolor{green!32} & \textbf{.017$^{\phantom{\dag}}\pm^{\phantom{\dag}}$.021} \cellcolor{green!40} & .028$^{\phantom{\dag}}\pm^{\phantom{\dag}}$.040 \cellcolor{green!6} & .042$^{\phantom{\dag}}\pm^{\phantom{\dag}}$.068 \cellcolor{red!33} & .028$^{\phantom{\dag}}\pm^{\phantom{\dag}}$.040 \cellcolor{green!6} & .044$^{\phantom{\dag}}\pm^{\phantom{\dag}}$.063 \cellcolor{red!40} \\
\bottomrule
\end{tabular}

%% file: tables/rKL_table.tex
\begin{tabular}{ccccc}
\toprule
\multicolumn{1}{c}{} & Naive@$k$ & CC & \add{PACC \footnotesize{(ours)}} & \add{KDEy \footnotesize{(ours)}} \\
\midrule
Geo. Location & .188$^{\dag}\pm^{\phantom{\dag}}$.255 \cellcolor{green!12} & .212$^{\phantom{\dag}}\pm^{\phantom{\dag}}$.246 \cellcolor{green!1} & .298$^{\phantom{\dag}}\pm^{\phantom{\dag}}$.241 \cellcolor{red!40} & \textbf{.132$^{\phantom{\dag}}\pm^{\phantom{\dag}}$.181} \cellcolor{green!40} \\
Gender & .305$^{\phantom{\dag}}\pm^{\phantom{\dag}}$.356 \cellcolor{red!40} & .068$^{\dag}\pm^{\phantom{\dag}}$.143 \cellcolor{green!30} & .064$^{\phantom{\dag}}\pm^{\phantom{\dag}}$.108 \cellcolor{green!32} & \textbf{.037$^{\phantom{\dag}}\pm^{\phantom{\dag}}$.060} \cellcolor{green!40} \\
Age of Topic & .213$^{\dag}\pm^{\phantom{\dag}}$.265 \cellcolor{red!3} & .173$^{\dag}\pm^{\phantom{\dag}}$.219 \cellcolor{green!17} & .285$^{\phantom{\dag}}\pm^{\phantom{\dag}}$.321 \cellcolor{red!40} & \textbf{.129$^{\phantom{\dag}}\pm^{\phantom{\dag}}$.158} \cellcolor{green!40} \\
\bottomrule
\end{tabular}

%% file: main.bbl
\vskip 0.2in


\begin{thebibliography}{}

\bibitem[\protect\BCAY{Adjaye{-}Gbewonyo, Bednarczyk, Davis,\ \BBA\ Omer}{Adjaye{-}Gbewonyo et~al.}{2014}]{adjaye2014using}
Adjaye{-}Gbewonyo, D., Bednarczyk, R.~A., Davis, R.~L., \BBA\ Omer, S.~B. \BBOP2014\BBCP.
\newblock \BBOQ Using the bayesian improved surname geocoding method (bisg) to create a working classification of race and ethnicity in a diverse managed care population: A validation study\BBCQ\
\newblock {\Bem Health Services Research}, {\Bem 49\/}(1), 268--283.

\bibitem[\protect\BCAY{Bella, Ferri, Hernández-Orallo,\ \BBA\ Ramírez-Quintana}{Bella et~al.}{2010}]{Bella:2010kx}
Bella, A., Ferri, C., Hernández-Orallo, J., \BBA\ Ramírez-Quintana, M.~J. \BBOP2010\BBCP.
\newblock \BBOQ Quantification via probability estimators\BBCQ\
\newblock In {\Bem Proceedings of the 11th IEEE International Conference on Data Mining (ICDM 2010)}, \BPGS\ 737--742, Sydney, {AU}.

\bibitem[\protect\BCAY{Biega, Gummadi,\ \BBA\ Weikum}{Biega et~al.}{2018}]{biega2018equity}
Biega, A.~J., Gummadi, K.~P., \BBA\ Weikum, G. \BBOP2018\BBCP.
\newblock \BBOQ Equity of attention: {Amortizing} individual fairness in rankings\BBCQ\
\newblock In {\Bem The 41st international acm sigir conference on research \& development in information retrieval}, \BPGS\ 405--414.

\bibitem[\protect\BCAY{Bogen, Rieke,\ \BBA\ Ahmed}{Bogen et~al.}{2020a}]{bogen2020awareness}
Bogen, M., Rieke, A., \BBA\ Ahmed, S. \BBOP2020a\BBCP.
\newblock \BBOQ Awareness in practice: {T}ensions in access to sensitive attribute data for antidiscrimination\BBCQ\
\newblock In {\Bem Proc, of FAccT}.

\bibitem[\protect\BCAY{Bogen, Rieke,\ \BBA\ Ahmed}{Bogen et~al.}{2020b}]{bogen2021:ap}
Bogen, M., Rieke, A., \BBA\ Ahmed, S. \BBOP2020b\BBCP.
\newblock \BBOQ Awareness in practice: {Tensions} in access to sensitive attribute data for antidiscrimination\BBCQ\
\newblock In {\Bem Proc. of the 3rd ACM Conference on Fairness, Accountability, and Transparency (FAT* 2020)}, \BPGS\ 492--500, Barcelona, ES.

\bibitem[\protect\BCAY{Bunse}{Bunse}{2022a}]{Bunse:2022oj}
Bunse, M. \BBOP2022a\BBCP.
\newblock \BBOQ On multi-class extensions of adjusted classify and count\BBCQ\
\newblock In {\Bem Proceedings of the 2nd International Workshop on Learning to Quantify (LQ 2022)}, \BPGS\ 43--50, Grenoble, IT.

\bibitem[\protect\BCAY{Bunse}{Bunse}{2022b}]{Bunse:2022ky}
Bunse, M. \BBOP2022b\BBCP.
\newblock \BBOQ Unification of algorithms for quantification and unfolding\BBCQ\
\newblock In {\Bem Proceedings of the Workshop on Machine Learning for Astroparticle Physics and Astronomy}, \BPGS\ 459--468.

\bibitem[\protect\BCAY{Celis, Huang, Keswani,\ \BBA\ Vishnoi}{Celis et~al.}{2021}]{celis2021fair}
Celis, L.~E., Huang, L., Keswani, V., \BBA\ Vishnoi, N.~K. \BBOP2021\BBCP.
\newblock \BBOQ Fair classification with noisy protected attributes: {A} framework with provable guarantees\BBCQ\
\newblock In {\Bem International Conference on Machine Learning}, \BPGS\ 1349--1361. PMLR.

\bibitem[\protect\BCAY{Chen\ \BBA\ Fang}{Chen\ \BBA\ Fang}{2023}]{chen2023learn}
Chen, F.\BBACOMMA\  \BBA\ Fang, H. \BBOP2023\BBCP.
\newblock \BBOQ Learn to be fair without labels: {A} distribution-based learning framework for fair ranking\BBCQ\
\newblock In {\Bem Proc. of SIGIR}, \BPGS\ 23--32.

\bibitem[\protect\BCAY{Chen, Kallus, Mao, Svacha,\ \BBA\ Udell}{Chen et~al.}{2019}]{chen2019fairness}
Chen, J., Kallus, N., Mao, X., Svacha, G., \BBA\ Udell, M. \BBOP2019\BBCP.
\newblock \BBOQ Fairness under unawareness: {Assessing} disparity when protected class is unobserved\BBCQ\
\newblock In {\Bem Proceedings of the Conference on Fairness, Accountability, and Transparency}, \BPGS\ 339--348.

\bibitem[\protect\BCAY{Chen, Ma, Hann{\'a}k,\ \BBA\ Wilson}{Chen et~al.}{2018}]{chen2018investigating}
Chen, L., Ma, R., Hann{\'a}k, A., \BBA\ Wilson, C. \BBOP2018\BBCP.
\newblock \BBOQ Investigating the impact of gender on rank in resume search engines\BBCQ\
\newblock In {\Bem Proceedings of the 2018 chi conference on human factors in computing systems}, \BPGS\ 1--14.

\bibitem[\protect\BCAY{Craswell, Zoeter, Taylor,\ \BBA\ Ramsey}{Craswell et~al.}{2008}]{craswell2008experimental}
Craswell, N., Zoeter, O., Taylor, M., \BBA\ Ramsey, B. \BBOP2008\BBCP.
\newblock \BBOQ An experimental comparison of click position-bias models\BBCQ\
\newblock In {\Bem Proc. of WSDM}.

\bibitem[\protect\BCAY{Diaz, Mitra, Ekstrand, Biega,\ \BBA\ Carterette}{Diaz et~al.}{2020}]{diaz2020evaluating}
Diaz, F., Mitra, B., Ekstrand, M.~D., Biega, A.~J., \BBA\ Carterette, B. \BBOP2020\BBCP.
\newblock \BBOQ Evaluating stochastic rankings with expected exposure\BBCQ\
\newblock In {\Bem Proc. of CIKM}.

\bibitem[\protect\BCAY{Ekstrand, Das, Burke, Diaz, et~al.}{Ekstrand et~al.}{2022a}]{ekstrand2022fairness}
Ekstrand, M.~D., Das, A., Burke, R., Diaz, F., et~al. \BBOP2022a\BBCP.
\newblock \BBOQ Fairness in information access systems\BBCQ\
\newblock {\Bem Foundations and Trends{\textregistered} in Information Retrieval}, {\Bem 16\/}(1-2), 1--177.

\bibitem[\protect\BCAY{Ekstrand, McDonald, Raj,\ \BBA\ Johnson}{Ekstrand et~al.}{2022b}]{trec-fair-ranking-2021}
Ekstrand, M.~D., McDonald, G., Raj, A., \BBA\ Johnson, I. \BBOP2022b\BBCP.
\newblock \BBOQ Overview of the trec 2021 fair ranking track\BBCQ\
\newblock In {\Bem Proc. of TREC}.

\bibitem[\protect\BCAY{Ekstrand, McDonald, Raj,\ \BBA\ Johnson}{Ekstrand et~al.}{2022c}]{trec-fair-ranking-2022}
Ekstrand, M.~D., McDonald, G., Raj, A., \BBA\ Johnson, I. \BBOP2022c\BBCP.
\newblock \BBOQ Overview of the {TREC 2022 Fair Ranking Track}\BBCQ\
\newblock In {\Bem Proc. of TREC}.

\bibitem[\protect\BCAY{Esuli, Fabris, Moreo,\ \BBA\ Sebastiani}{Esuli et~al.}{2023}]{Esuli2023}
Esuli, A., Fabris, A., Moreo, A., \BBA\ Sebastiani, F. \BBOP2023\BBCP.
\newblock {\Bem Learning to quantify}.
\newblock Springer Nature, Cham, CH.

\bibitem[\protect\BCAY{{European Commission}}{{European Commission}}{2024}]{EUAIACT2024}
{European Commission} \BBOP2024\BBCP.
\newblock \BBOQ Regulation on harmonized rules on artificial intelligence (artificial intelligence act)\BBCQ\
\newblock Official Journal of the European Union.
\newblock Accessed: 2024-08-06.

\bibitem[\protect\BCAY{Fabris, Esuli, Moreo,\ \BBA\ Sebastiani}{Fabris et~al.}{2023}]{fabris2023measuring}
Fabris, A., Esuli, A., Moreo, A., \BBA\ Sebastiani, F. \BBOP2023\BBCP.
\newblock \BBOQ Measuring fairness under unawareness of sensitive attributes: {A} quantification-based approach\BBCQ\
\newblock {\Bem Journal of Artificial Intelligence Research}, {\Bem 76}, 1117--1180.

\bibitem[\protect\BCAY{Forman}{Forman}{2005}]{Forman:2005fk}
Forman, G. \BBOP2005\BBCP.
\newblock \BBOQ Counting positives accurately despite inaccurate classification\BBCQ\
\newblock In {\Bem Proceedings of the 16th European Conference on Machine Learning (ECML 2005)}, \BPGS\ 564--575, Porto, {PT}.

\bibitem[\protect\BCAY{Friedler, Scheidegger,\ \BBA\ Venkatasubramanian}{Friedler et~al.}{2021}]{friedler2021possibility}
Friedler, S.~A., Scheidegger, C., \BBA\ Venkatasubramanian, S. \BBOP2021\BBCP.
\newblock \BBOQ The (im)possibility of fairness: {D}ifferent value systems require different mechanisms for fair decision making\BBCQ\
\newblock {\Bem Communications of the ACM}, {\Bem 64\/}(4), 136--143.

\bibitem[\protect\BCAY{Geyik, Ambler,\ \BBA\ Kenthapadi}{Geyik et~al.}{2019}]{geyik2019fairness}
Geyik, S.~C., Ambler, S., \BBA\ Kenthapadi, K. \BBOP2019\BBCP.
\newblock \BBOQ Fairness-aware ranking in search and recommendation systems with application to {Linkedin} talent search\BBCQ\
\newblock In {\Bem Proc. of KDD}.

\bibitem[\protect\BCAY{Ghazimatin, Kleindessner, Russell, Abedjan,\ \BBA\ Golebiowski}{Ghazimatin et~al.}{2022}]{ghazimatin2022measuring}
Ghazimatin, A., Kleindessner, M., Russell, C., Abedjan, Z., \BBA\ Golebiowski, J. \BBOP2022\BBCP.
\newblock \BBOQ Measuring fairness of rankings under noisy sensitive information\BBCQ\
\newblock In {\Bem Proc. of FAccT}.

\bibitem[\protect\BCAY{Ghosh, Dutt,\ \BBA\ Wilson}{Ghosh et~al.}{2021}]{ghosh2021fair}
Ghosh, A., Dutt, R., \BBA\ Wilson, C. \BBOP2021\BBCP.
\newblock \BBOQ When fair ranking meets uncertain inference\BBCQ\
\newblock In {\Bem Proc. of SIGIR}.

\bibitem[\protect\BCAY{Ghosh, Kvitca,\ \BBA\ Wilson}{Ghosh et~al.}{2023}]{ghosh2023fair}
Ghosh, A., Kvitca, P., \BBA\ Wilson, C. \BBOP2023\BBCP.
\newblock \BBOQ When fair classification meets noisy protected attributes\BBCQ\
\newblock In {\Bem Proceedings of the 2023 AAAI/ACM Conference on AI, Ethics, and Society}, \BPGS\ 679--690.

\bibitem[\protect\BCAY{Gonz{\'a}lez, Moreo,\ \BBA\ Sebastiani}{Gonz{\'a}lez et~al.}{2024}]{gonzalez2024binary}
Gonz{\'a}lez, P., Moreo, A., \BBA\ Sebastiani, F. \BBOP2024\BBCP.
\newblock \BBOQ Binary quantification and dataset shift: an experimental investigation\BBCQ.

\bibitem[\protect\BCAY{Hardt, Price,\ \BBA\ Srebro}{Hardt et~al.}{2016}]{hardt2016equality}
Hardt, M., Price, E., \BBA\ Srebro, N. \BBOP2016\BBCP.
\newblock \BBOQ Equality of opportunity in supervised learning\BBCQ\
\newblock {\Bem Proc. of NeurIPS}, {\Bem 29}.

\bibitem[\protect\BCAY{Heuss, Sarvi,\ \BBA\ de~Rijke}{Heuss et~al.}{2022}]{heuss2022fairness}
Heuss, M., Sarvi, F., \BBA\ de~Rijke, M. \BBOP2022\BBCP.
\newblock \BBOQ Fairness of exposure in light of incomplete exposure estimation\BBCQ\
\newblock In {\Bem Proc. of SIGIR}.

\bibitem[\protect\BCAY{Holstein, Wortman~Vaughan, Daum{\'e}~III, Dudik,\ \BBA\ Wallach}{Holstein et~al.}{2019}]{holstein2019improving}
Holstein, K., Wortman~Vaughan, J., Daum{\'e}~III, H., Dudik, M., \BBA\ Wallach, H. \BBOP2019\BBCP.
\newblock \BBOQ Improving fairness in machine learning systems: {What do industry practitioners need?}\BBCQ\
\newblock In {\Bem Proc. of CHI}.

\bibitem[\protect\BCAY{Jaenich, McDonald,\ \BBA\ Ounis}{Jaenich et~al.}{2023}]{jaenich2023colbert}
Jaenich, T., McDonald, G., \BBA\ Ounis, I. \BBOP2023\BBCP.
\newblock \BBOQ {ColBERT-FairPRF: Towards} fair pseudo-relevance feedback in dense retrieval\BBCQ\
\newblock In {\Bem European Conference on Information Retrieval}, \BPGS\ 457--465. Springer.

\bibitem[\protect\BCAY{Jaenich, McDonald,\ \BBA\ Ounis}{Jaenich et~al.}{2024}]{jaenich2024fairness}
Jaenich, T., McDonald, G., \BBA\ Ounis, I. \BBOP2024\BBCP.
\newblock \BBOQ Fairness-aware exposure allocation via adaptive reranking\BBCQ\
\newblock In {\Bem Proceedings of the 47th International ACM SIGIR Conference on Research and Development in Information Retrieval}, \BPGS\ 1504--1513.

\bibitem[\protect\BCAY{K{\i}rnap, Diaz, Biega, Ekstrand, Carterette,\ \BBA\ Yilmaz}{K{\i}rnap et~al.}{2021}]{kirnap2021estimation}
K{\i}rnap, {\"O}., Diaz, F., Biega, A., Ekstrand, M., Carterette, B., \BBA\ Yilmaz, E. \BBOP2021\BBCP.
\newblock \BBOQ Estimation of fair ranking metrics with incomplete judgments\BBCQ\
\newblock In {\Bem Proc. of The Web Conference}.

\bibitem[\protect\BCAY{Kopeinik, Mara, Ratz, Krieg, Schedl,\ \BBA\ Rekabsaz}{Kopeinik et~al.}{2023}]{Kopeinik2023MaleNurse}
Kopeinik, S., Mara, M., Ratz, L., Krieg, K., Schedl, M., \BBA\ Rekabsaz, N. \BBOP2023\BBCP.
\newblock \BBOQ Show me a "male nurse"! how gender bias is reflected in the query formulation of search engine users\BBCQ\
\newblock In {\Bem Proc. of (CHI)}.

\bibitem[\protect\BCAY{Kuhlman, Gerych,\ \BBA\ Rundensteiner}{Kuhlman et~al.}{2021}]{kuhlman2021measuring}
Kuhlman, C., Gerych, W., \BBA\ Rundensteiner, E. \BBOP2021\BBCP.
\newblock \BBOQ Measuring group advantage: {A} comparative study of fair ranking metrics\BBCQ\
\newblock In {\Bem Proceedings of the 2021 AAAI/ACM Conference on AI, Ethics, and Society}, \BPGS\ 674--682.

\bibitem[\protect\BCAY{{LinkedIn}}{{LinkedIn}}{2024}]{linkedin2024settings}
{LinkedIn} \BBOP2024\BBCP.
\newblock \BBOQ Linkedin settings\BBCQ\
\newblock New York City Council Legislation.
\newblock Accessed: 2024-08-06.

\bibitem[\protect\BCAY{Lipton, Wang,\ \BBA\ Smola}{Lipton et~al.}{2018}]{Lipton:2018fj}
Lipton, Z.~C., Wang, Y., \BBA\ Smola, A.~J. \BBOP2018\BBCP.
\newblock \BBOQ Detecting and correcting for label shift with black box predictors\BBCQ\
\newblock In {\Bem Proceedings of the 35th International Conference on Machine Learning (ICML 2018)}, \BPGS\ 3128--3136, Stockholm, {SE}.

\bibitem[\protect\BCAY{Macdonald, Tonellotto, MacAvaney,\ \BBA\ Ounis}{Macdonald et~al.}{2021}]{macdonald2021pyterrier}
Macdonald, C., Tonellotto, N., MacAvaney, S., \BBA\ Ounis, I. \BBOP2021\BBCP.
\newblock \BBOQ {PyTerrier: Declarative} experimentation in {Python} from {BM25} to dense retrieval\BBCQ\
\newblock In {\Bem Proc. of CIKM}.

\bibitem[\protect\BCAY{Moreo, Esuli,\ \BBA\ Sebastiani}{Moreo et~al.}{2021}]{moreo2021quapy}
Moreo, A., Esuli, A., \BBA\ Sebastiani, F. \BBOP2021\BBCP.
\newblock \BBOQ {QuaPy: A Python}-based framework for quantification\BBCQ\
\newblock In {\Bem Proceedings of the 30th ACM International Conference on Knowledge Management (CIKM 2021)}, \BPGS\ 4534--4543, Gold Coast, AU.

\bibitem[\protect\BCAY{Moreo, González,\ \BBA\ del Coz}{Moreo et~al.}{2024}]{Moreo:2024KDEy}
Moreo, A., González, P., \BBA\ del Coz, J.~J. \BBOP2024\BBCP.
\newblock \BBOQ Kernel density estimation for multiclass quantification\BBCQ.

\bibitem[\protect\BCAY{Morik, Singh, Hong,\ \BBA\ Joachims}{Morik et~al.}{2020}]{morik2020controlling}
Morik, M., Singh, A., Hong, J., \BBA\ Joachims, T. \BBOP2020\BBCP.
\newblock \BBOQ Controlling fairness and bias in dynamic learning-to-rank\BBCQ\
\newblock In {\Bem Proc. of SIGIR}.

\bibitem[\protect\BCAY{Mozannar, Ohannessian,\ \BBA\ Srebro}{Mozannar et~al.}{2020}]{mozannar2020fair}
Mozannar, H., Ohannessian, M., \BBA\ Srebro, N. \BBOP2020\BBCP.
\newblock \BBOQ Fair learning with private demographic data\BBCQ\
\newblock In {\Bem International Conference on Machine Learning}, \BPGS\ 7066--7075. PMLR.

\bibitem[\protect\BCAY{{New York City Council}}{{New York City Council}}{2021}]{NYC2021LL144}
{New York City Council} \BBOP2021\BBCP.
\newblock \BBOQ Local law 144 of 2021\BBCQ\
\newblock New York City Council Legislation.
\newblock Accessed: 2024-08-06.

\bibitem[\protect\BCAY{Pedreschi, Ruggieri,\ \BBA\ Turini}{Pedreschi et~al.}{2008}]{pedreshi2008discrimination}
Pedreschi, D., Ruggieri, S., \BBA\ Turini, F. \BBOP2008\BBCP.
\newblock \BBOQ Discrimination-aware data mining\BBCQ\
\newblock In {\Bem Proc. of KDD}.

\bibitem[\protect\BCAY{Raj\ \BBA\ Ekstrand}{Raj\ \BBA\ Ekstrand}{2022}]{raj2022measuring}
Raj, A.\BBACOMMA\  \BBA\ Ekstrand, M.~D. \BBOP2022\BBCP.
\newblock \BBOQ Measuring fairness in ranked results: {An} analytical and empirical comparison\BBCQ\
\newblock In {\Bem Proc. of SIGIR}.

\bibitem[\protect\BCAY{Robertson, Walker, Jones, Hancock-Beaulieu, Gatford, et~al.}{Robertson et~al.}{1995}]{robertson1995okapi}
Robertson, S.~E., Walker, S., Jones, S., Hancock-Beaulieu, M.~M., Gatford, M., et~al. \BBOP1995\BBCP.
\newblock \BBOQ Okapi at {TREC-3}\BBCQ\
\newblock {\Bem NIST Special Publication Sp}, {\Bem 109}, 109.

\bibitem[\protect\BCAY{Sapiezynski, Zeng, E~Robertson, Mislove,\ \BBA\ Wilson}{Sapiezynski et~al.}{2019}]{sapiezynski2019quantifying}
Sapiezynski, P., Zeng, W., E~Robertson, R., Mislove, A., \BBA\ Wilson, C. \BBOP2019\BBCP.
\newblock \BBOQ Quantifying the impact of user attentionon fair group representation in ranked lists\BBCQ\
\newblock In {\Bem Proc. of Companion Proceedings of WWW}.

\bibitem[\protect\BCAY{Sch{\"{o}}lkopf, Janzing, Peters, Sgouritsa, Zhang,\ \BBA\ Mooij}{Sch{\"{o}}lkopf et~al.}{2012}]{Scholkopf:2012je}
Sch{\"{o}}lkopf, B., Janzing, D., Peters, J., Sgouritsa, E., Zhang, K., \BBA\ Mooij, J.~M. \BBOP2012\BBCP.
\newblock \BBOQ On causal and anticausal learning\BBCQ\
\newblock In {\Bem Proceedings of the 29th International Conference on Machine Learning (ICML 2012)}, Edinburgh, {UK}.

\bibitem[\protect\BCAY{Schumacher, Strohmaier,\ \BBA\ Lemmerich}{Schumacher et~al.}{2023}]{schumacher2023comparative}
Schumacher, T., Strohmaier, M., \BBA\ Lemmerich, F. \BBOP2023\BBCP.
\newblock \BBOQ A comparative evaluation of quantification methods\BBCQ.

\bibitem[\protect\BCAY{Sebastiani}{Sebastiani}{2020}]{Sebastiani:2020qf}
Sebastiani, F. \BBOP2020\BBCP.
\newblock \BBOQ Evaluation measures for quantification: {A}n axiomatic approach\BBCQ\
\newblock {\Bem Information Retrieval Journal}, {\Bem 23\/}(3), 255--288.

\bibitem[\protect\BCAY{Simson, Fabris,\ \BBA\ Kern}{Simson et~al.}{2024}]{fabris2024lazy}
Simson, J., Fabris, A., \BBA\ Kern, C. \BBOP2024\BBCP.
\newblock \BBOQ Lazy data practices harm fairness research\BBCQ\
\newblock In {\Bem The 2024 {ACM} Conference on Fairness, Accountability, and Transparency, FAccT 2024, Rio de Janeiro, Brazil, June 3-6, 2024}, \BPGS\ 642--659. {ACM}.

\bibitem[\protect\BCAY{Singh\ \BBA\ Joachims}{Singh\ \BBA\ Joachims}{2018}]{singh2018fairness}
Singh, A.\BBACOMMA\  \BBA\ Joachims, T. \BBOP2018\BBCP.
\newblock \BBOQ Fairness of exposure in rankings\BBCQ\
\newblock In {\Bem Proc. of KDD}.

\bibitem[\protect\BCAY{Storkey}{Storkey}{2009}]{Storkey:2009lp}
Storkey, A. \BBOP2009\BBCP.
\newblock \BBOQ When training and test sets are different: {C}haracterizing learning transfer\BBCQ\
\newblock In Quiñonero-Candela, J., Sugiyama, M., Schwaighofer, A., \BBA\ Lawrence, N.~D.\BEDS, {\Bem Dataset shift in machine learning}, \BPGS\ 3--28. The {MIT} Press, Cambridge, {US}.

\bibitem[\protect\BCAY{Wang, Guo, Narasimhan, Cotter, Gupta,\ \BBA\ Jordan}{Wang et~al.}{2020}]{wang2020robust}
Wang, S., Guo, W., Narasimhan, H., Cotter, A., Gupta, M., \BBA\ Jordan, M. \BBOP2020\BBCP.
\newblock \BBOQ Robust optimization for fairness with noisy protected groups\BBCQ\
\newblock {\Bem Advances in Neural Information Processing systems}, {\Bem 33}, 5190--5203.

\bibitem[\protect\BCAY{Wilson, Ghosh, Jiang, Mislove, Baker, Szary, Trindel,\ \BBA\ Polli}{Wilson et~al.}{2021}]{wilson2021building}
Wilson, C., Ghosh, A., Jiang, S., Mislove, A., Baker, L.~J., Szary, J., Trindel, K., \BBA\ Polli, F. \BBOP2021\BBCP.
\newblock \BBOQ Building and auditing fair algorithms: {A} case study in candidate screening\BBCQ\
\newblock In Elish, M.~C., Isaac, W., \BBA\ Zemel, R.~S.\BEDS, {\Bem FAccT '21: 2021 {ACM} Conference on Fairness, Accountability, and Transparency, Virtual Event / Toronto, Canada, March 3-10, 2021}, \BPGS\ 666--677. {ACM}.

\bibitem[\protect\BCAY{Yang, J\"{a}nich, Mayfield,\ \BBA\ Lawrie}{Yang et~al.}{2024}]{yang2024language}
Yang, E., J\"{a}nich, T., Mayfield, J., \BBA\ Lawrie, D. \BBOP2024\BBCP.
\newblock \BBOQ Language fairness in multilingual information retrieval\BBCQ\
\newblock In {\Bem Proc. of Companion Proceedings of SIGIR}.

\bibitem[\protect\BCAY{Yang\ \BBA\ Stoyanovich}{Yang\ \BBA\ Stoyanovich}{2017}]{yang2017measuring}
Yang, K.\BBACOMMA\  \BBA\ Stoyanovich, J. \BBOP2017\BBCP.
\newblock \BBOQ Measuring fairness in ranked outputs\BBCQ\
\newblock In {\Bem Proc. of SSDBM}.

\bibitem[\protect\BCAY{Zehlike, Bonchi, Castillo, Hajian, Megahed,\ \BBA\ Baeza-Yates}{Zehlike et~al.}{2017}]{zehlike2017fa}
Zehlike, M., Bonchi, F., Castillo, C., Hajian, S., Megahed, M., \BBA\ Baeza-Yates, R. \BBOP2017\BBCP.
\newblock \BBOQ {Fa*ir: A} fair top-k ranking algorithm\BBCQ\
\newblock In {\Bem Proc. of CIKM}.

\bibitem[\protect\BCAY{Zehlike, Yang,\ \BBA\ Stoyanovich}{Zehlike et~al.}{2022}]{zehlike2022fairness}
Zehlike, M., Yang, K., \BBA\ Stoyanovich, J. \BBOP2022\BBCP.
\newblock \BBOQ Fairness in ranking, part {I}: {Score}-based ranking\BBCQ\
\newblock {\Bem ACM Computing Surveys}, {\Bem 55\/}(6), 1--36.

\end{thebibliography}
